\documentclass{IEEEtran}
\usepackage{cite}
\usepackage{amsmath,amssymb,amsfonts}
\usepackage{algorithmic}
\usepackage{graphicx}
\usepackage{textcomp}

\usepackage{xcolor}
\usepackage{circuitikz}
\usepackage{multirow}
\usepackage{subcaption}
\usepackage{bm}
\usepackage{float}

\usepackage{booktabs}

\definecolor{purple}{RGB}{87,80,155}
\definecolor{blue}{RGB}{0,114,178}
\definecolor{green}{RGB}{0,158,115}
\definecolor{pink}{RGB}{204,121,167}
\definecolor{lightgray}{RGB}{245,245,245}

\def\BibTeX{{\rm B\kern-.05em{\sc i\kern-.025em b}\kern-.08em
    T\kern-.1667em\lower.7ex\hbox{E}\kern-.125emX}}
\begin{document}
\title{Lossless Compression via Chained Lightweight Neural Predictors with Information Inheritance}
\author{Yuriy Kim, Evgeny Belyaev 
\thanks{Received 14 April 2026.
(Corresponding author: Yuriy Kim)}
\thanks{Yuriy Kim and Evgeny Belyaev are with the ITMO University, Saint-Petersburg, Russia (e-mail: ylkim@itmo.ru; eabelyaev@itmo.ru).}
}
\maketitle

\begin{abstract}
This paper is dedicated to lossless data compression with probability estimation using neural networks. First, we propose a probability estimation architecture based on a
chain of neural predictors, so that each unit of the chain is defined as a neural network
with the minimum possible number of weights, which is sufficient for efficient compression of data generated by Markov sources of a given order. We show that this architecture allows us to minimize the overall number of weights participating in the probability estimation process depending on the statistical properties of the input data. Second, in order to improve
compression efficiency, we introduce an information inheritance mechanism, where the probability estimate obtained by a low-order unit is used at the
next higher-order unit. Experimental results show that the proposed lossless data 
compressor equipped with the chained probability estimation architecture provides
compression ratios close to the state-of-the-art PAC compressor.
At the same time, it outperforms PAC by a factor of 1.2 to 6.3 in encoding throughput
and by a factor of 2.8 to 12.3 in decoding throughput on a consumer GPU.
\end{abstract}


\section{Introduction}
\label{sec:introduction}
The amount of digital data generated worldwide continues to grow at an unprecedented pace, and lossless compression remains a critical technology for modern information systems. In 2024 alone, approximately 22 zettabytes of data were generated globally~\cite{idc_datasphere_2025_2029}, and this amount is expected to exceed 146~zettabytes by 2029, corresponding to an annual growth rate of 46\%. This growing category includes a wide range of data sources, such as software, emails and office documents, business application data, collaborative content, web infrastructure logs, metadata, and data produced by large-scale analytics pipelines and artificial intelligence or machine learning systems. The scale and diversity of this data make efficient universal lossless compression essential for sustainable storage, transmission, and processing infrastructures.

Classical general-purpose lossless compression algorithms, including dictionary-based methods such as LZ77 and LZ78~\cite{lz77} and context-modeling approaches such as PPMD~\cite{cleary2003ppmd}, have demonstrated remarkable robustness and efficiency over decades of use. These algorithms are valued for their high throughput, low memory footprint, and deterministic decoding, which make them well-suited for deployment in large-scale systems. However, their compression performance is fundamentally limited by manually designed statistical models and restricted context representations. As a result, while these methods are fast, they often fail to fully exploit the complex, high-order dependencies present in data streams.

Recent advances in machine learning have sparked growing interest in lossless compression based on neural networks, which replaces handcrafted probabilistic models with learned predictors. By estimating the conditional probability distribution of the next symbol given its context and combining these predictions with arithmetic coding~\cite{witten1987arithmetic}, neural compressors can, in principle, approach tighter entropy bounds for complex sources. Empirically, such methods have demonstrated compression ratios that surpass the classical algorithms on a wide range of datasets, highlighting their potential as a new generation of universal compressors.

From a probability modeling perspective, neural compression methods can be classified~\cite{goyal2021dzip} into the following categories:
\begin{enumerate}
\item In \textit{static} schemes, a model is trained off-line on external data and shared by both encoder and decoder, i.e.
the model weights are not embedded into the compressed file header. This approach is efficient
only if the model can efficiently estimate  
probability models for a wide range of different sources.
Therefore, in practice, this approach is mainly applicable to large language models (LLMs), such as Llama 2 and Chinchilla~\cite{deletang2023language}.
\item \textit{Adaptive} methods initialize the encoder and decoder with identical random models that are updated 
synchronously at the encoder and decoder. As a result, the neural network can learn the probability model of the input data, and there is no need to transmit its weights to the decoder. However, it requires training the network on the decoder side, which decreases its throughput.  
Currently, well-known examples of adaptive schemes include  
TRACE~\cite{mao2022trace} and PAC~\cite{mao2023pac}
universal compressors. TRACE utilizes a single-layer Transformer architecture together with a byte-grouping scheme, while
 PAC achieves more efficient compression due to a multilayer perceptron (MLP) based predictor and reduced repetitive processing of previous symbols in context modeling (the duplicated-processing problem). 

\item \textit{Semi-adaptive} methods  occupy an intermediate position between the static and the adaptive ones: the neural model is trained specifically on the input sequence, potentially over multiple passes. Then its weights and other parameters are transmitted in the compressed file header. This introduces an explicit trade-off between the model prediction ability and 
the number of bits required for the header. At the same time, it avoids the training procedure on the decoder side, increasing its 
throughput. For example, operating in bootstrap mode, the
DZip~\cite{goyal2021dzip} compressor first trains the recurrent neural network (RNN) on the sequence before compression, and 
then stores the learned weights in the compressed file header and uses them for encoding and decoding. 
\end{enumerate}

Excluding extremely slow LLM approaches with throughput less than 0.1--1KB/s on consumer GPUs, currently PAC is the state-of-the-art universal compressor that provides higher compression ratios compared to the classical compression algorithms
and compressors with neural probability modeling, such as TRACE and DZip. However, its encoding and decoding throughput reaches
34--40 KB/s on NVIDIA GeForce RTX 4060 Ti, which is still too slow for modern needs. Therefore, the main goal of this paper is to find an alternative neural probability modeling architecture that provides much higher encoding and decoding throughput while keeping compression efficiency. In this work, we were motivated by the following observations:
\begin{enumerate}
\item Statistical properties of different types of
data vary significantly. For example, 
the probability distribution for the current symbol in 
text files depends on hundreds of previous symbols, while
symbols from random data do not depend on each other. However, 
PAC uses a single neural architecture for any kind of
input data. This means that it wastes computational resources
compressing "simple" sources which could be efficiently compressed by 
neural architectures with a much smaller number of weights.
\item The adaptive probability modeling used in PAC 
means that the decoding process should involve computationally expensive backpropagation for network training. However, it is preferable to avoid network training,
which increases decoding throughput.
\end{enumerate}

Taking the above observations into account, in this paper we propose an efficient lossless compression architecture constructed from chained lightweight neural predictors. This architecture adaptively minimizes the number of weights participating in the probability estimation process according to the statistical properties of the input 
data\footnote{Some preliminary experiments in this direction
have been done by the authors in~\cite{kim25}}.
The main contributions of this paper are summarized as follows:
\begin{enumerate}
    \item We propose a probability estimation architecture via chained neural predictors. The first unit of this chain
    is a neural network that provides strong compression performance for Markov sources of order $s=1$ using the minimum number of weights. The second unit is a neural network
    selected in the same manner for Markov sources of order $s=2$
    and so on. This architecture allows us to select the minimum number of units for given data, achieving good compression ratios while minimizing the overall number of weights participating in the probability estimation process.
    \item Inspired by context-based probability modeling in the classic PPMII~\cite{PPMII} compressor, we introduce the information inheritance mechanism, so that
    the probability estimate obtained by a lower-order unit is used at the next higher-order unit. We show that this mechanism increases the accuracy of the estimation compared with the probability estimation via a single unit.
    \item We introduce an efficient lossless data compressor based on the chained architecture with data-driven tokenization, semi-adaptive probability estimation by the unit chain, and arithmetic encoding. The semi-adaptive nature of the proposed compressor allows us to avoid the use of backpropagation on the decoder side, thereby increasing
    its throughput.
    \item Experiments show that the proposed compressor provides compression ratios
    very close to PAC with the encoding throughput from 43 to 244 KB/s and 
    the decoding throughput from 98 to 429 KB/s on GPU NVIDIA GeForce RTX 4060 Ti. 
\end{enumerate}

The rest of the paper is organized as follows. Section~\ref{sec:method}-A
describes the proposed lightweight neural predictors architecture construction
via Markov source modeling, Section~\ref{sec:method}-B introduces the information inheritance approach for the considered chained architecture, 
Section~\ref{sec:method}-C describes the proposed data-driven tokenization, Section~\ref{sec:method}-D shows how the proposed chained architecture is used for
lossless data compression, and Section~\ref{sec:method}-E shows how the number of weights participating in the probability modeling can be minimized. The experimental results are presented in Section~\ref{sec:experiments}, while Section~\ref{sec:conclusion} concludes the presented results.

\section{Proposed method}
\label{sec:method}
\subsection{Lightweight Neural Predictors Architecture Construction via Markov source modeling}
\label{sec:markovmodeling}

\begin{figure*}[t]
    \centering
    \begin{tabular}{ccc}
        \begin{subfigure}[b]{0.3\textwidth}
            \includegraphics[width=\textwidth]{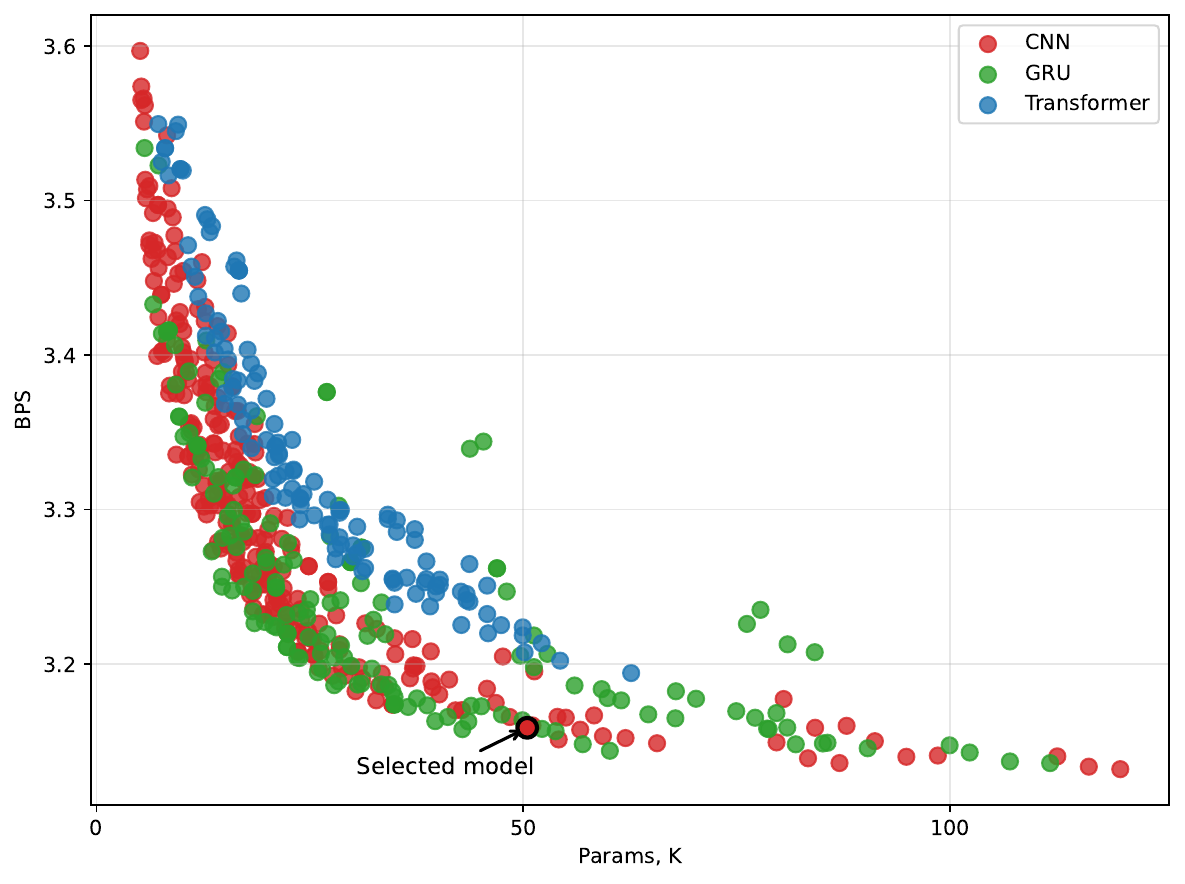}
            \caption{$s_2=2$}
        \end{subfigure} &
        \begin{subfigure}[b]{0.3\textwidth}
            \includegraphics[width=\textwidth]{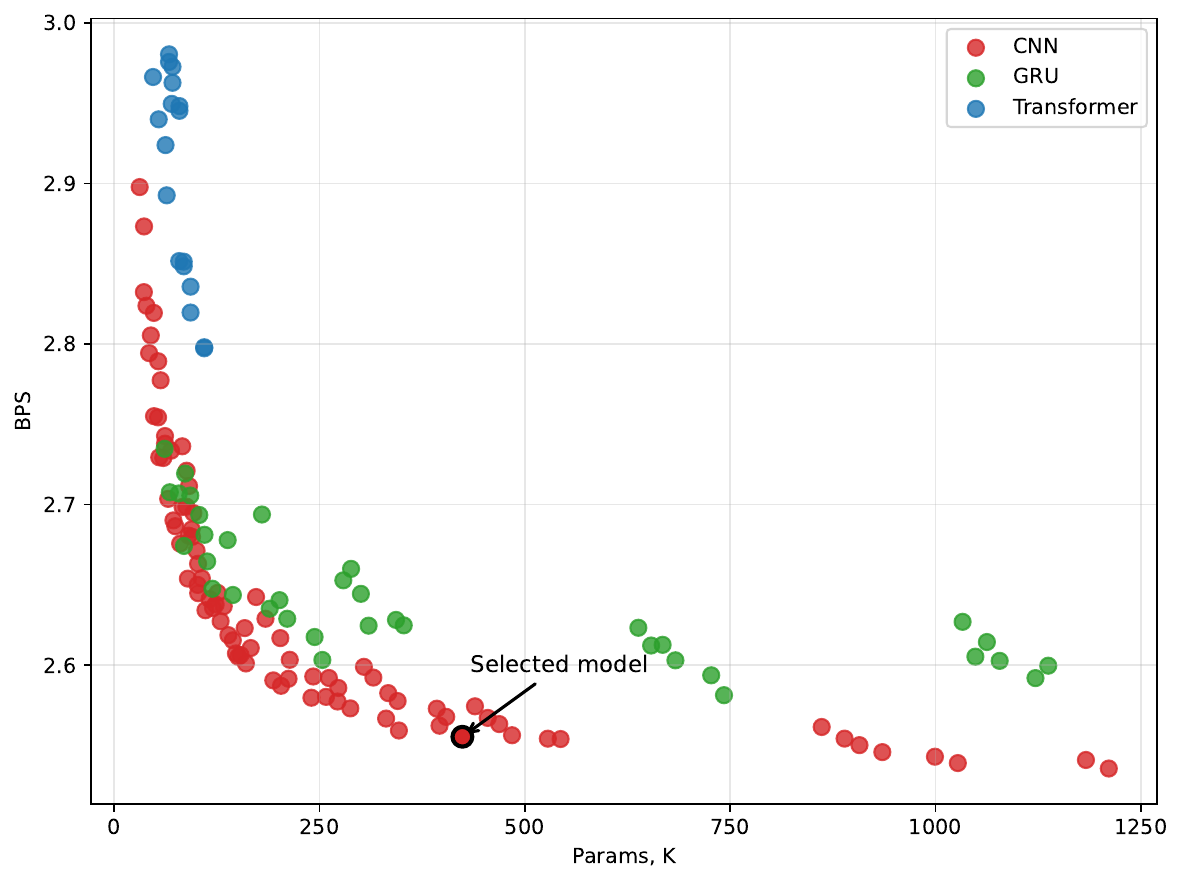}
            \caption{$s_3=3$}
        \end{subfigure} &
        \begin{subfigure}[b]{0.3\textwidth}
            \includegraphics[width=\textwidth]{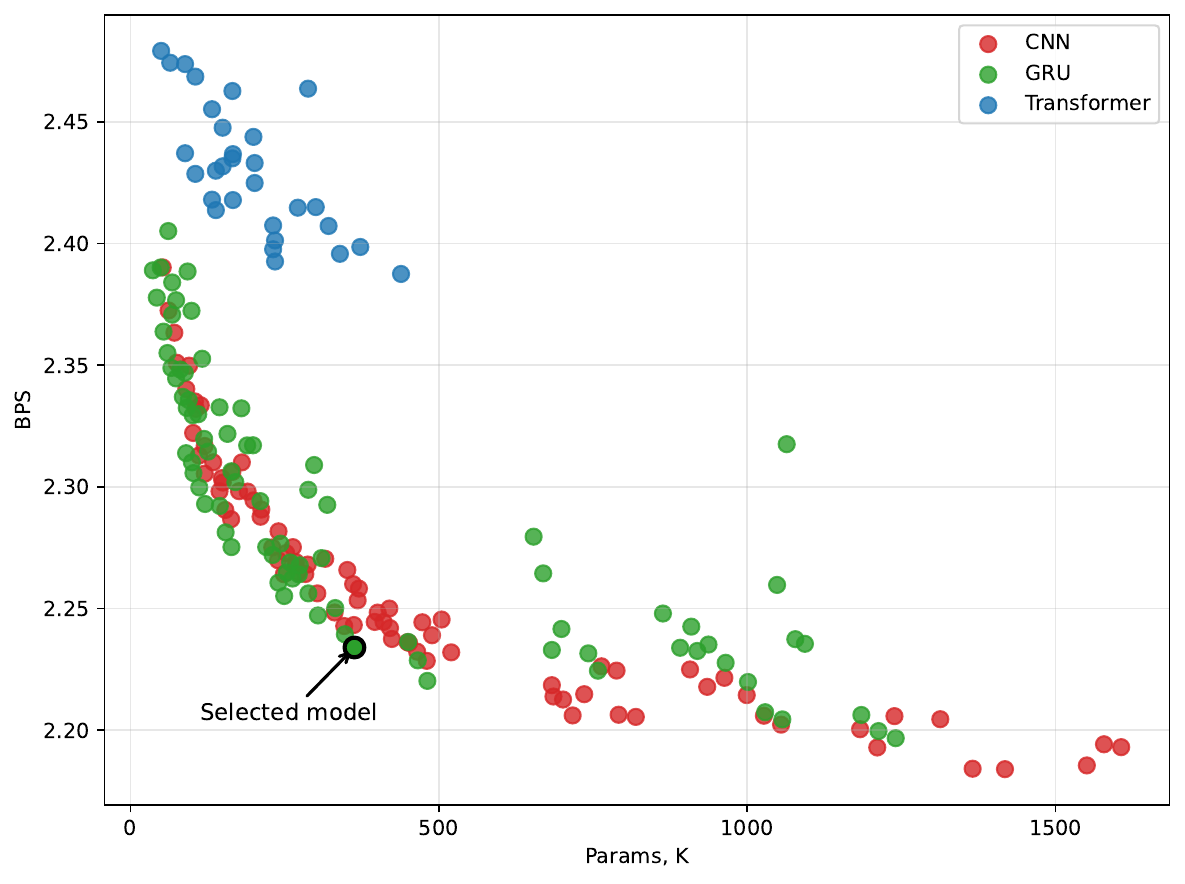}
            \caption{$s_4=4$}
        \end{subfigure}
    \end{tabular}
    \caption{\label{fig:arch_search} The proposed neural architecture search visualization}
\end{figure*}

In this paper, we propose to construct neural architectures
by modeling real-life data with a Markov source.
Let us define $\bm x^j_i$ as a short notation for the sequence $x_i,x_{i+1},...,x_{j-1},x_j$ and assume that the sequence $\bm x^N_1 = \{x_1,x_2,...,x_N\}$, $x_i \in \bm X$ is modeled by a Markov source of order $s$, i.e. $p(x_i|\bm x^{i-1}_1)=p(x_i|\bm x^{i-1}_{i-s})$. Then the proposed approach is performed in the following
stages:
\begin{enumerate}
\item For $\bm x^N_1$, each conditional probability $p(x_i|\bm x^{i-1}_{i-s})$ for a Markov source of order $s$ is estimated as
\begin{equation}
\hat p(x_i|\bm x^{i-1}_{i-s}) = \frac{n(\bm x^{i}_{i-s})}{n(\bm x^{i-1}_{i-s})},
\end{equation}
where $n(\bm x^j_i)$ is the number of occurrences (frequency) of the subsequence $\bm x^j_i$ in $\bm x^N_1$ and $\hat p(x_i|\bm x^{i-1}_{i-s})=0$, if $n(\bm x^{i-1}_{i-s})=0$. 

\item  For the Markov source with order $s_1 \in \bm S$ 
we generate a realization $\bm y_1^M$, so that
$p(y_i|\bm y^{i-1}_{i-s_1})\approx \hat p(x_i|\bm x^{i-1}_{i-s_1})$.
Then we search for the
type of  neural architecture and its parameters $\theta_1 \in \bm \Theta$, such that
\begin{equation}
\left\{
\begin{array}{l}
\theta_1 = \arg \displaystyle\min_{\theta \in \bm \Theta} \mathcal C(\theta)\\
\mathcal{L} (\theta_1,s_1) \leq \kappa M\cdot H({s_1}),
\end{array}
\right.
\end{equation}
where $\mathcal C(\theta)$ is the complexity of the model,  
\begin{equation}
\mathcal{L} (\theta_1,s_1) = -\sum_{i=1}^{M}
\log_2 \hat p(y_i|\bm y^{i-1}_{i-s_1},\theta_1)
\end{equation}
is an estimate of the number of bits that can be obtained by the model, and $\hat p(y_i|\bm y^{i-1}_{i-s_1},\theta_1)$
is the probability estimate of symbol $y_i$ for context 
$\bm y^{i-1}_{i-s_1}$ given by the model $\theta_1$, and
\begin{equation}
H({s_1}) = \frac{1}{M}\left(-\log_2p(\bm y^{s_1}_1)-\displaystyle\sum^M_{i=s_1+1}\log_2 p(y_i|\bm y^{i-1}_{i-s_1})\right)
\end{equation}
is the bits per symbol bound for compression of the realization $\bm y^M_1$.

\item Similarly, for the Markov source of order $s_2 \in \bm S$ 
we generate a realization $\bm y_1^M$, so that
$p(y_i|\bm y^{i-1}_{i-s_2})\approx \hat p(x_i|\bm x^{i-1}_{i-s_2})$. Then, we search for the
neural architecture $\theta_2 \in \bm \Theta$, such that
\begin{equation}
\left\{
\begin{array}{l}
\theta_2 = \arg \displaystyle\min_{\theta \in \bm \Theta} \mathcal C(\theta)\\
\mathcal{L} (\theta_2,s_2) \leq \kappa M\cdot H({s_2}).
\end{array}
\right.
\end{equation}
The main difference from the previous stage is that we apply the information inheritance approach, where the probability estimate 
$\hat p(y_i|\bm y^{i-1}_{i-s_1},\theta_1)$ obtained by the  model $\theta_1$ is also used by the model $\theta_2$:
\begin{equation}
\mathcal{L} (\theta_2,s_2)=-\sum_{i=1}^{M}
\log_2 \hat p(y_i|\bm y^{i-1}_{i-s_2},\theta_2,\hat p(y_i|\bm y^{i-1}_{i-s_1},\theta_1)).
\end{equation}
\item For orders $s_3,s_4,...,s_L$, we use the same idea as in the previous stage, where the probability estimate obtained by model $\theta_i$ is used by model $\theta_{i+1}$.
\end{enumerate}

\begin{figure*}[!ht]
\centering
\resizebox{1\textwidth}{!}{%
\begin{circuitikz}
\tikzstyle{every node}=[font=\large]
\draw [fill=green!30, rounded corners=5pt] (3.75,18) rectangle  node {\large $Emb (16)$} (7.25,17);
\draw [fill=blue!40,rounded corners=5pt] (3.75,14.5) rectangle  node {\large $MLP(128) $} (7.25,13.5);
\draw [fill=purple!40,rounded corners=5pt] (3.75,16.25) rectangle  node {\large $Conv1D(2, 128) $} (7.25,15.25);
\draw [fill=green!30,rounded corners=5pt] (-0.75,16) rectangle  node {\large $Emb(8)$} (2.75,15.25);
\draw [fill=blue!40,rounded corners=5pt] (-0.75,14.25) rectangle  node {\large $MLP(16) $} (2.75,13.5);
\draw [fill=green!30,rounded corners=5pt] (8.25,18) rectangle  node {\large $Emb(32)$} (11.75,17);
\draw [fill=blue!40,rounded corners=5pt] (8.25,14.5) rectangle  node {\large $MLP(512) $} (11.75,13.5);
\draw [fill=purple!40,rounded corners=5pt] (8.25,16.25) rectangle  node {\large $Conv1D(3, 512) $} (11.75,15.25);
\draw [fill=green!30,rounded corners=5pt] (12.75,18) rectangle  node {\large $Emb(48)$} (16.25,17);
\draw [fill=blue!40,rounded corners=5pt] (12.75,14.5) rectangle  node {\large $MLP(256) $} (16.25,13.5);
\draw [fill=pink!40,rounded corners=5pt] (12.75,16.25) rectangle  node {\large $GRU(256) $} (16.25,15.25);
\draw [fill=green!30,rounded corners=5pt] (17.25,18) rectangle  node {\large $Emb(32)$} (20.75,17);
\draw [fill=blue!40,rounded corners=5pt] (17.25,14.5) rectangle  node {\large $MLP(512) $} (20.75,13.5);
\draw [fill=green!30,rounded corners=5pt] (21.75,18) rectangle  node {\large $Emb(48)$} (25.25,17);
\draw [fill=blue!40,rounded corners=5pt] (21.75,14.5) rectangle  node {\large $MLP(512) $} (25.25,13.5);
\draw [fill=pink!40,rounded corners=5pt] (21.75,16.25) rectangle  node {\large $GRU(256) $} (25.25,15.25);
\draw [fill=purple!40,rounded corners=5pt] (17.25,16.5) rectangle (20.75,15);
\node [font=\large] at (19,15.4) {$\times2$};
\node [font=\large] at (19,15.9) {$Conv1D(3, 256) $};
\draw [fill=lightgray] (5.5,11.5) circle (0.5cm) node {\large $+$} ;
\draw [fill=lightgray] (10,11.5) circle (0.5cm) node {\large $+$} ;
\draw [fill=lightgray] (14.5,11.5) circle (0.5cm) node {\large $+$} ;
\draw [fill=lightgray] (19,11.5) circle (0.5cm) node {\large $+$} ;
\draw [fill=lightgray] (23.5,11.5) circle (0.5cm) node {\large $+$} ;
\draw [->, >=Stealth] (19,17) -- (19,16.5);
\draw [->, >=Stealth] (19,15) -- (19,14.5);
\node [font=\large] at (5,12.25) {$\alpha_2$};
\node [font=\large] at (4.5,11.25) {$\beta_2$};
\node [font=\large] at (9.5,12.25) {$\alpha_3$};
\node [font=\large] at (1.25,13) {$l'_1$};
\node [font=\large] at (9,11.25) {$\beta_3$};
\node [font=\large] at (6,13) {$l'_2$};
\node [font=\large] at (10.5,13) {$l'_3$};
\node [font=\large] at (6.5,11.75) {$l_2$};
\draw [->, >=Stealth] (5.5,17) -- (5.5,16.25);
\draw [->, >=Stealth] (5.5,19) -- (5.5,18);
\draw [->, >=Stealth] (5.5,15.25) -- (5.5,14.5);
\draw [->, >=Stealth] (5.5,13.5) -- (5.5,12);
\draw [->, >=Stealth] (1,11.5) -- (5,11.5);
\draw [->, >=Stealth] (6,11.5) -- (9.5,11.5);
\draw [->, >=Stealth] (1,15.25) -- (1,14.25);
\draw [->, >=Stealth] (1,19.5) -- (1,16);
\draw [short] (1,13.5) -- (1,11.5);
\draw [->, >=Stealth] (10,17) -- (10,16.25);
\draw [->, >=Stealth] (10,19) -- (10,18);
\draw [->, >=Stealth] (10,15.25) -- (10,14.5);
\draw [->, >=Stealth] (10,13.5) -- (10,12);
\draw [->, >=Stealth] (10.5,11.5) -- (14,11.5);
\draw [->, >=Stealth] (14.5,17) -- (14.5,16.25);
\draw [->, >=Stealth] (14.5,19) -- (14.5,18);
\draw [->, >=Stealth] (14.5,15.25) -- (14.5,14.5);
\draw [->, >=Stealth] (14.5,13.5) -- (14.5,12);
\node [font=\large] at (14,12.25) {$\alpha_4$};
\node [font=\large] at (13.5,11.25) {$\beta_4$};
\node [font=\large] at (15,13) {$l'_4$};
\draw [->, >=Stealth] (15,11.5) -- (18.5,11.5);
\draw [->, >=Stealth] (19,19) -- (19,18);
\draw [->, >=Stealth] (19,13.5) -- (19,12);
\node [font=\large] at (18.5,12.25) {$\alpha_5$};
\node [font=\large] at (18,11.25) {$\beta_5$};
\node [font=\large] at (19.5,13) {$l'_5$};
\draw [->, >=Stealth] (19.5,11.5) -- (23,11.5);
\draw [->, >=Stealth] (23.5,17) -- (23.5,16.25);
\draw [->, >=Stealth] (23.5,19) -- (23.5,18);
\draw [->, >=Stealth] (23.5,15.25) -- (23.5,14.5);
\draw [->, >=Stealth] (23.5,13.5) -- (23.5,12);
\node [font=\large] at (23,12.25) {$\alpha_6$};
\node [font=\large] at (22.5,11.25) {$\beta_6$};
\node [font=\large] at (24,13) {$l'_6$};
\draw [->, >=Stealth] (24,11.5) -- (25.25,11.5);
\node [font=\large] at (11,11.75) {$l_3$};
\node [font=\large] at (15.5,11.75) {$l_4$};
\node [font=\large] at (20,11.75) {$l_5$};
\node [font=\large] at (24.5,11.75) {$l_6$};
\draw [short] (1,19) -- (23.5,19);
\node [font=\large] at (1,19.75) {$x_1,x_2,...,x_N$};
\node [font=\large] at (24.75,18.5) {$\bm x^{i-16}_{i-1}$};
\node [font=\large] at (20.25,18.5) {$\bm x^{i-8}_{i-1}$};
\node [font=\large] at (15.75,18.5) {$\bm x^{i-4}_{i-1}$};
\node [font=\large] at (11.25,18.5) {$\bm x^{i-3}_{i-1}$};
\node [font=\large] at (6.5,18.5) {$\bm x^{i-2}_{i-1}$};
\node [font=\large] at (1.5,18.5) {$x_{i-1}$};
\end{circuitikz}
}%

\caption{\label{fig:cascade_scheme} The proposed architecture of chained neural predictors with information inheritance}
\end{figure*}
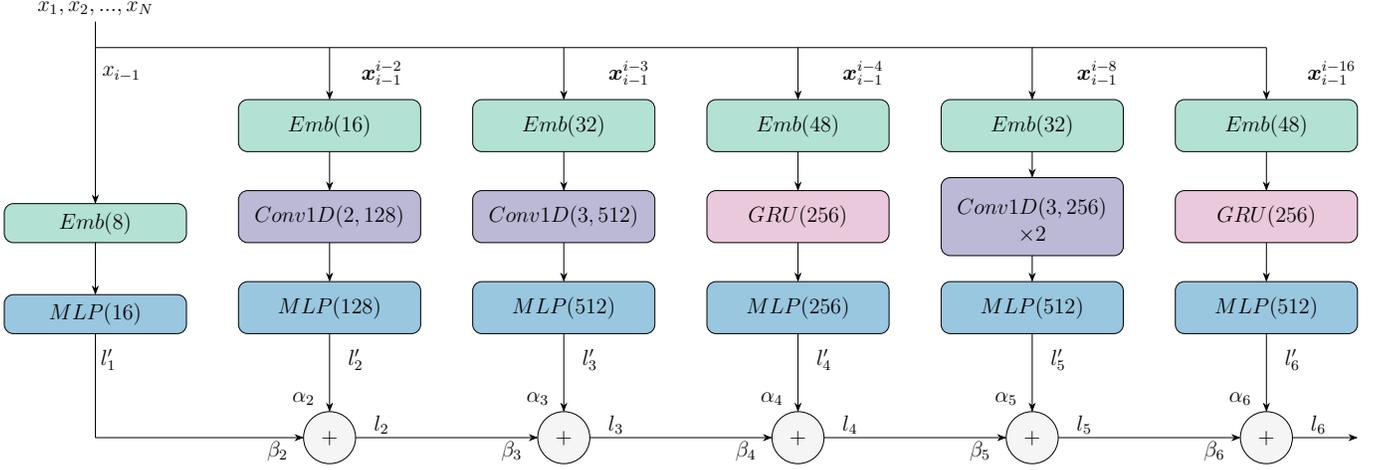

As an example, Figure~\ref{fig:arch_search} shows a visualization of the search for neural architectures for $s_2=2$, $s_3=3$, and $s_4=4$, given $s_1=1$.
The considered neural architecture classes include:
\begin{enumerate}
    \item Multilayer Perceptron (MLP) (only for $s_1=1)$;
    \item Convolutional Neural Network (CNN);
    \item Gated Recurrent Unit (GRU);
    \item Transformer.
\end{enumerate}
For each architecture type, we changed the following parameters:
\begin{enumerate}
    \item Number of layers;
    \item Embedding dimension;
    \item Hidden dimension;
    \item Dimension of fully connected layers;
    \item Kernel size (for CNN);
    \item Number of attention heads (for Transformer).
\end{enumerate}

Here we used the enwik9 dataset~\cite{mahoney2011large} with $N=10^9$ as an input real-life sequence. In order to find the best neural architecture for each source order, we performed a grid search for $\kappa=1.05$ and $M=10^8$.
Each point in Figure~\ref{fig:arch_search} corresponds to bits per symbol (BPS) and compression time for a candidate model from the architecture search, while the highlighted point indicates the model selected for that order. To keep the search computationally feasible and the models comparable, we imposed an upper bound of 1M parameters. One can see that, surprisingly, different architecture types
provide the best results at different source orders. For example, CNN provides the best results for orders $s_2=2$, $s_3=3$, while GRU is better for $s_4=4$.

\subsection{The proposed architecture of chained neural predictors with information inheritance}

Without loss of generality, in this paper we selected $L=6$ with $\bm S \in \{1,2,3,4,8,16\}$.
The first four architectures are obtained for $s_1=1$, $s_2=2$, $s_3=3$, and $s_4=4$, as described in Section~\ref{sec:markovmodeling}. For higher orders $s \in \{8,16\}$, the bound $H(s_j)$ cannot be estimated directly due to limitations in memory size needed for context storage, as well as unreliable estimates obtained by frequency counting. Therefore, for higher orders we estimate $H(s_j)$ as
\begin{equation}
H(s_j) \approx \min_{\theta \in \bm \Theta} \mathcal L (\theta,s_j).
\end{equation}

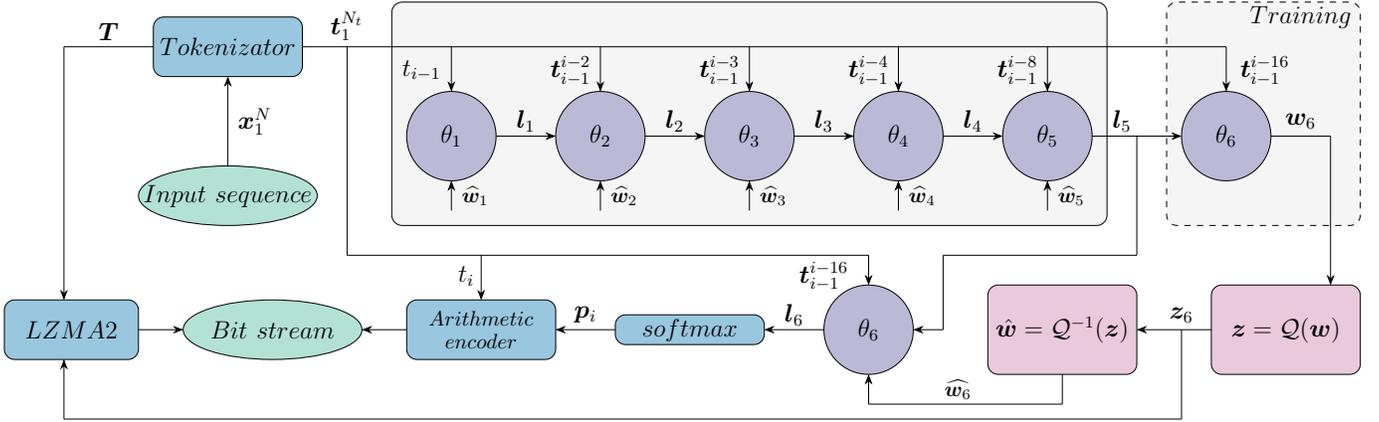
\begin{figure*}[!ht]
\centering
\resizebox{1\textwidth}{!}{%
\begin{circuitikz}
\tikzstyle{every node}=[font=\large]
\tikzstyle{every node}=[font=\large]
\draw [fill=lightgray,rounded corners=5pt] (14,11.5) rectangle (26,7.75);
\draw [fill=lightgray,rounded corners=5pt] [ dashed] (27,11.5) rectangle  (30.25,7.75);
\draw [fill=pink!40,rounded corners=5pt] (24,6.75) rectangle  node {\large $\hat{\bm w} = \mathcal Q^{-1}(\bm z)$} (26.5,5.25);
\draw [fill=pink!40,rounded corners=5pt] (27.75,6.75) rectangle  node {\large $\bm z = \mathcal Q(\bm w)$} (30.25,5.25);
\draw [fill=blue!40,rounded corners=5pt] (10,11.25) rectangle  node {\large $Tokenizator$} (12.5,10.25);
\draw [fill=purple!40] (15,9.25) circle (0.75cm) node {\large $\theta_1$} ;
\draw [fill=purple!40] (17.5,9.25) circle (0.75cm) node {\large $\theta_2$} ;
\draw [fill=purple!40] (20,9.25) circle (0.75cm) node {\large $\theta_3$} ;
\draw [fill=purple!40] (22.5,9.25) circle (0.75cm) node {\large $\theta_4$} ;
\draw [fill=purple!40] (25,9.25) circle (0.75cm) node {\large $\theta_5$} ;
\draw [fill=purple!40] (28,9.25) circle (0.75cm) node {\large $\theta_6$} ;
\draw [short] (12.5,10.75) -- (28,10.75);
\draw [short] (19.25,12) -- (19.25,12);
\draw [short] (20,12) -- (20,12);
\draw [->, >=Stealth] (15,10.75) -- (15,10);
\draw [->, >=Stealth] (17.5,10.75) -- (17.5,10);
\draw [->, >=Stealth] (20,10.75) -- (20,10);
\draw [->, >=Stealth] (22.5,10.75) -- (22.5,10);
\draw [->, >=Stealth] (25,10.75) -- (25,10);
\draw [->, >=Stealth] (28,10.75) -- (28,10);
\draw [->, >=Stealth] (15.75,9.25) -- (16.75,9.25);
\draw [->, >=Stealth] (18.25,9.25) -- (19.25,9.25);
\draw [->, >=Stealth] (20.75,9.25) -- (21.75,9.25);
\draw [->, >=Stealth] (23.25,9.25) -- (24.25,9.25);
\draw [->, >=Stealth] (25.75,9.25) -- (27.25,9.25);
\draw [fill=blue!40,rounded corners=5pt] (14.25,6.5) rectangle  node {} (16.75,5.5);
\node [font=\normalsize] at (15.5,6.2) {$Arithmetic$};
\node [font=\normalsize] at (15.5,5.8) {$encoder$};
\draw [->, >=Stealth] (27.75,6) -- (26.5,6);
\draw [->, >=Stealth] (17.75,6) -- (16.75,6);
\draw [short] (13.25,7.25) -- (22,7.25);
\draw [short] (27.25,4.5) -- (8.5,4.5);
\node [font=\large] at (11.7,9.5) {$\bm x_1^N$};
\node [font=\large] at (13.25,11.1) {$\bm t_1^{N_t}$};
\node [font=\large] at (29.25,9.5) {$\bm w_6$};
\node [font=\large] at (27.25,6.25) {$\bm z_6$};
\node [font=\normalsize] at (23.5,5) {$\widehat{\bm w_6}$};
\node [font=\large] at (16.25,9.5) {$\bm l_1$};
\node [font=\large] at (18.75,9.5) {$\bm l_2$};
\node [font=\large] at (21.25,9.5) {$\bm l_3$};
\node [font=\large] at (23.75,9.5) {$\bm l_4$};
\node [font=\large] at (26.25,9.5) {$\bm l_5$};
\node [font=\large] at (17.25,6.25) {$\bm p_i$};
\node [font=\large] at (21.25,6.9) {$\bm t^{i-16}_{i-1}$};
\node [font=\large] at (15.25,6.9) {$t_i$};
\draw [fill=blue!40,rounded corners=5pt] (9.75,6.5) rectangle  node {\large $LZMA2$} (7.5,5.5);
\draw [->, >=Stealth] (8.5,4.5) -- (8.5,5.5);
\draw [fill=green!30] (12,6) ellipse (1.5cm and 0.5cm) node {\large $Bit\ stream$} ;
\draw  (7.5,13.25) circle (0cm);
\draw [fill=green!30] (11.25,8.25) ellipse (1.5cm and 0.5cm) node {\large $Input\ sequence$} ;
\draw [->, >=Stealth] (15,8) -- (15,8.5);
\draw [->, >=Stealth] (17.5,8) -- (17.5,8.5);
\draw [->, >=Stealth] (20,8) -- (20,8.5);
\draw [->, >=Stealth] (22.5,8) -- (22.5,8.5);
\draw [->, >=Stealth] (25,8) -- (25,8.5);
\node [font=\normalsize] at (15.4,8.25) {$\widehat{\bm w}_1$};
\node [font=\large] at (29.25,11.25) {$Training$};
\node [font=\normalsize] at (17.9,8.25) {$\widehat{\bm w}_2$};
\node [font=\normalsize] at (25.4,8.25) {$\widehat{\bm w}_5$};
\node [font=\normalsize] at (22.9,8.25) {$\widehat{\bm w}_4$};
\node [font=\normalsize] at (20.4,8.25) {$\widehat{\bm w}_3$};
\draw [fill=purple!40] (22,6) circle (0.75cm) node {\large $\theta_6$} ;
\draw [short] (26.5,9.25) -- (26.5,7.25);
\draw [short] (26.5,7.25) -- (23.25,7.25);
\draw [fill=blue!40,rounded corners=5pt] (17.75,6.25) rectangle  node {\large $softmax$} (20.25,5.75);
\draw [->, >=Stealth] (22,7.25) -- (22,6.75);
\draw [short] (28.75,9.25) -- (29.75,9.25);
\draw [short] (25.25,5.25) -- (25.25,4.75);
\draw [short] (25.25,4.75) -- (22,4.75);
\draw [->, >=Stealth] (22,4.75) -- (22,5.25);
\draw [short] (23.25,7.25) -- (23.25,6);
\draw [->, >=Stealth] (23.25,6) -- (22.75,6);
\draw [->, >=Stealth] (21.25,6) -- (20.25,6);
\node [font=\large] at (20.75,6.25) {$\bm l_6$};
\draw [short] (10,10.75) -- (8.5,10.75);
\draw [->, >=Stealth] (8.5,10.75) -- (8.5,6.5);
\draw [->, >=Stealth] (9.75,6) -- (10.5,6);
\draw [->, >=Stealth] (14.25,6) -- (13.5,6);
\draw [->, >=Stealth] (11.25,8.75) -- (11.25,10.25);
\draw [short] (27.25,4.5) -- (27.25,6);
\node [font=\large] at (24.5,10.35) {$\bm t^{i-8}_{i-1}$};
\node [font=\large] at (14.5,10.35) {$t_{i-1}$};
\node [font=\large] at (17,10.35) {$\bm t^{i-2}_{i-1}$};
\node [font=\large] at (19.5,10.35) {$\bm t^{i-3}_{i-1}$};
\node [font=\large] at (22,10.35) {$\bm t^{i-4}_{i-1}$};
\node [font=\large] at (28.65,10.35) {$\bm t^{i-16}_{i-1}$};
\node [font=\large] at (9.25,11) {$\bm T$};
\draw [->, >=Stealth] (29.75,9.25) -- (29.75,6.75);
\draw [short] (13.25,10.75) -- (13.25,7.25);
\draw [->, >=Stealth] (15.5,7.25) -- (15.5,6.5);
\end{circuitikz}
}%

\caption{\label{fig:general_scheme} Training of neural predictor $\theta_6$ and full encoding pipeline }
\end{figure*}

Figure~\ref{fig:cascade_scheme} shows the resulting architecture with six chained units. In the figure, $Emb(d)$ denotes the embedding layer, where $d$ is the embedding dimension; $Conv1D(k,c)$ denotes a one-dimensional convolutional block, where $k$ is the kernel size and $c$ is the number of output channels; $GRU(h)$ is a gated recurrent unit block with hidden size $h$; and $MLP(d)$ is a multilayer perceptron head with hidden size $d$ that produces logits. 

First of all, each predictor for order $s_i$ has an MLP layer of the corresponding size in order to compute the output logit value $l'_i$.
Second, as mentioned in earlier context-based probability modeling based on 
Prediction by Partial Matching (PPM), the similarity of distribution functions in
parent and child contexts can be used to improve prediction efficiency
via information inheritance~\cite{PPMII}. In this paper, we propose to use the same
idea, but within the scope of the machine learning approach. The information inheritance at order $s_i \neq s_1$ is proposed to realize via linear weighing of the logits $l'_{i+1}$ and $l'_i$ as
\begin{equation}
l_i=\alpha_i\cdot l'_{i}+\beta_i\cdot l'_{i-1},
\end{equation}
where $\alpha_i$ and $\beta_i$ are trainable parameters. Here,
$\alpha_i=1$, $\beta_i=0$ means that the information inheritance is disabled.
Finally, 
the conditional probabilities for each predictor is computed as
\begin{equation}
\label{probcomputing}
\hat{p}(x_i \mid \bm{x}^{i-1}_{i-s_j}, \theta_j) = 
\begin{cases}
\operatorname{softmax}(l'_j), & \text{if } j = 1, \\
\operatorname{softmax}(l_j), & \text{if } j \neq 1.
\end{cases}
\end{equation}
One of the main advantages of the proposed chained architecture presented in Figure~\ref{fig:cascade_scheme}, equipped with 
the conditional probabilities computation via (\ref{probcomputing}) is that it allows
to disable high-order units if faster encoding or decoding is needed, or
when high-order modeling does not give any compression advantage.
In other words, the source-order modeling provided by this architecture 
can be aligned with the order of the input data minimizing the number of weights participating in the probability estimation process. 

\subsection{Alphabet tokenization}
We use Byte Pair Encoding (BPE)~\cite{gage1994bpe} to map the input byte sequence $\bm x_1^N$ to the token sequence $\bm t_1^{N_k}$, where $N_k\le N$. Starting from the initial alphabet $\bm T_0=\bm X$, and sequence $\bm t_1^{N_0}=\bm x_1^N$, BPE greedily applies  the following $K$ merge operations. At step $k$, the 
alphabet $\bm T_k$ is computed as $\bm T_k = \{\bm T_{k-1}, t_k\}$, so that
\begin{equation}
\label{greedyrule}
t_k = \arg\max_{a,b\in \bm T_{k-1}} n(ab),
\end{equation}
where $n(ab)$ is the frequency of pair $ab$ in the current tokenized sequence $\bm t_1^{N_{k-1}}$. 

On the one hand, the greedy rule~(\ref{greedyrule}) reduces the resulting token sequence $\bm t_1^{N_k}$, decreasing the total number of calls to both the arithmetic encoder and decoder.
On the other hand, the growth of the alphabet size of $\bm T_k$ leads to an increase in complexity due to the higher number of parameters in the $Emb()$ and $MLP()$ layers of each neural probability predictor, and the number of operations needed for cumulative probability computation in arithmetic coding. Therefore, in this paper we propose
to select the alphabet $\bm T_j$, $j=1,...,K$, so that
$j = \displaystyle\min_{\Delta_i\leq\bar\Delta} i$, where $\Delta_i=N_{i-1}-N_i$, 
and $\bar\Delta = \frac{1}{K}\displaystyle\sum^K_{i=1} \Delta_i$, i.e.
we select the shortest alphabet for which the sequence-length reduction is lower than the average one.

\begin{figure}[t]
    \centering
    \begin{tabular}{cc}
        \begin{subfigure}[b]{0.45\columnwidth}
            \includegraphics[width=\textwidth]{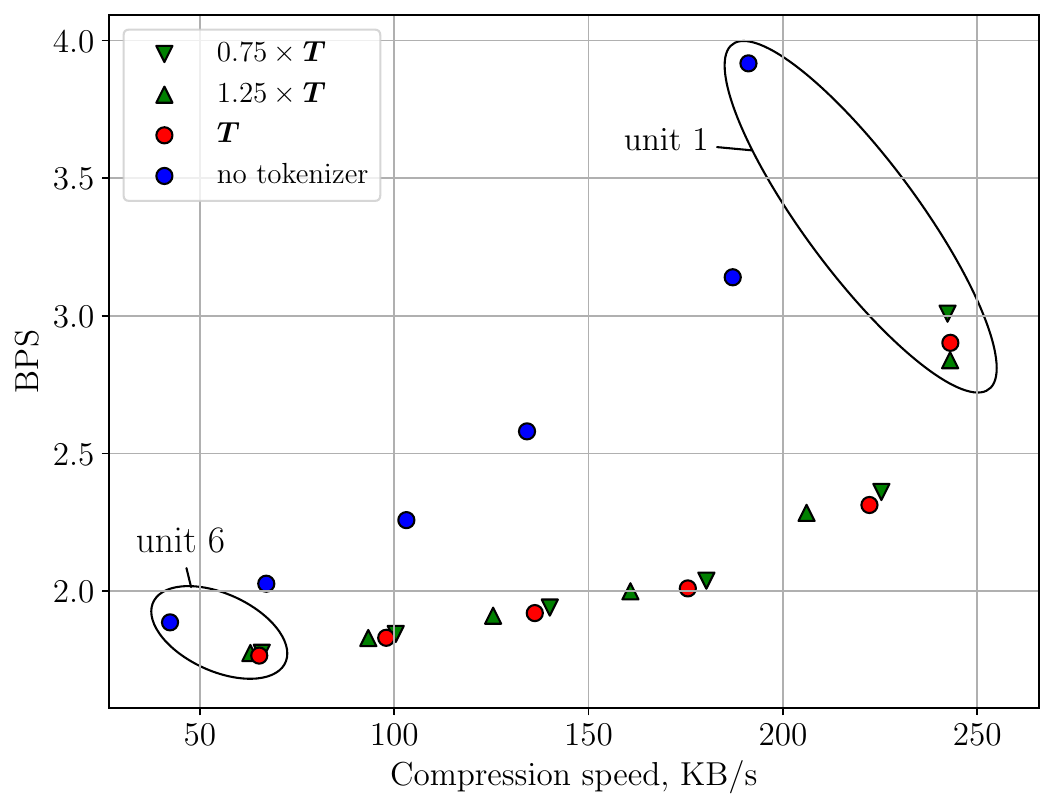}
            \caption{}
        \end{subfigure} &
        \begin{subfigure}[b]{0.45\columnwidth}
            \includegraphics[width=\textwidth]{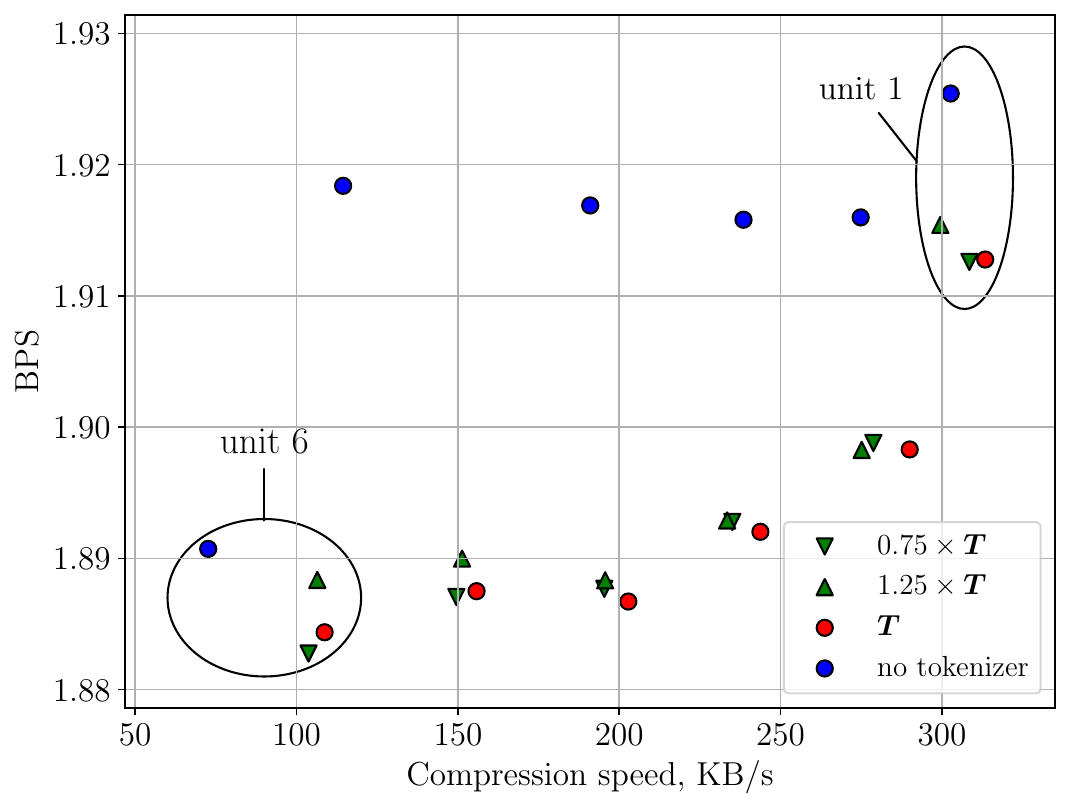}
            \caption{}
        \end{subfigure}
    \end{tabular}
    \caption{Compression performance for models with and without tokenization, as well as with alphabet sizes equal to $0.75\times$, $1.0\times$, and $1.25\times$ the selected alphabet $\bm T$ for (a) text dataset (enwik8) and (b) genome dataset (GaGa).}
    \label{fig:ablation_tokenizer}
\end{figure}

Figure~\ref{fig:ablation_tokenizer} compares the compression performance of the proposed model without tokenization, with the selected tokenizer, and with reduced and enlarged alphabets corresponding to $0.75\times$ and $1.25\times$ the selected alphabet $\bm T$. We can see that on average selected tokenizer provides the best result. At the same time, the impact of tokenization is dataset-dependent: for some datasets the gain is moderate, whereas for others it is more pronounced. The clearest benefit is observed for text and genomic data, where tokenization more effectively captures recurring multi-symbol patterns and more strongly reduces the token-sequence length. This reduction also leads to higher processing speed, since fewer tokens need to be processed by the neural predictors and the arithmetic coder.

\subsection{Encoding by the chained neural predictors}
Without loss of generality, let us assume that the tokenized alphabet $\bm T$ and the quantized weights 
$\hat{\bm w}_1$, ..., $\hat{\bm w}_5$, for the neural predictors $\theta_1$, ..., $\theta_5$
are already selected and stored in the header of the compressed bit stream. Then let us consider the encoding of the input sequence of tokens $t_1,...,t_{N_t}$ by the proposed chained neural predictors (see Figure~\ref{fig:general_scheme}).

First, we train the neural predictor $\theta_6$ to find its weights $\bm w_6$ by minimizing the negative log-likelihood measured in bits using the probabilities defined in (\ref{probcomputing}). The training is performed in two stages. During the first $E_1$ epochs, the predictor is pretrained without information inheritance, i.e., with $\alpha_6=1$ and $\beta_6=0$. During the next $E_2$ epochs, information inheritance is enabled and all model parameters are jointly optimized. The pretraining allows the current predictor to first learn a meaningful standalone representation, after which the inherited information can be incorporated more effectively. For a batch $\mathcal B$, the training objective is

\begin{equation}
\mathcal L_{\mathcal B}(\theta_6,\bm w_6)= -\sum_{j\in\mathcal B}\log_2 p(t_j\mid \bm t_{j-16}^{j-1},\theta_6,\bm w_6),
\end{equation}
where $t_j$ is the target token and $\bm t_{j-16}^{j-1}$ is its context. Let $N_t$ denote the number of tokens in the input sequence and let the validation subset contain $N_{\mathrm{v}}=\rho N_t$ tokens, where $\rho \in (0,1]$ is the validation coefficient. For early stopping, the loss on this validation subset is evaluated every $K_\mathrm{v}$ training iterations. Training is stopped if this loss does not improve for $P$ consecutive checks.

Second, for the trained neural predictor $\theta_6$, we perform post-training compression of model parameters $\bm w_6$  combining optimizer-moment-guided pruning as in ExCP~\cite{Li2024ExCP}  and vector quantization.
The pruning is applied element-wise as
\begin{equation}
\label{pruning}
w_6(i) \leftarrow
\begin{cases}
w_6(i), & \text{if } |w_6(i)| \ge r_w, \\
0, & \text{otherwise},
\end{cases}
\end{equation}
where threshold $r_w$ is defined as
\begin{equation}
r_w = \frac{\gamma}{\sqrt{m_t}}\,\mathrm{median}(\bm w_6),
\end{equation}
and $m_t$ denotes the second-order moment of the optimizer, and $\gamma$ is an amplifying parameter.

The pruned weights $\bm w_6$ are flattened and divided into non-overlapped vectors $\bm v_1,\bm v_2,...$ of size $V$.
Then each vector $\bm v_t$ is approximated by vector $\bm c_k \in \mathcal C$, $|\mathcal C|=2^b$, via vector quantization, where index $k$ of centroid $\bm c_k$ in the code book~$\mathcal C$ is selected as
\begin{equation}
k = \arg\min_{i} \lVert\mathbf v_t-\mathbf c_i\rVert_2.
\end{equation}
The code book $\mathcal C$ is learned using k-means algorithm.

\begin{table}[t]
\caption{Description of the datasets used in the experiments}
\centering
\begin{tabular}{|c|c|c|p{0.4\linewidth}|}
\hline
\textbf{Name} & \textbf{Type} & \textbf{Size} & \textbf{Description} \\
\hline
Backup100 & \multirow{2}{*}{Heterogeneous} & 100MB 
& \multirow{2}{=}{Random extract bytes from the disk backup~\cite{mao2022trace}} \\
\cline{1-1} \cline{3-3}
Backup &  & 1GB & \\
\hline
Enwik8 & \multirow{3}{*}{Text} & 100MB 
& \multirow{2}{=}{English Wikipedia dump (2006)~\cite{mahoney2011large}} \\
\cline{1-1} \cline{3-3}
Enwik9 &  & 1GB & \\
\cline{1-1} \cline{3-4}
Books &  & 100MB
& BookCorpus~\cite{vaswani2017attention} \\
\hline
Image100 & \multirow{2}{*}{Image} & 100MB 
& \multirow{2}{=}{ImageNet~\cite{deng2009imagenet}} \\
\cline{1-1} \cline{3-3}
Image &  & 1.2GB 
&  \\
\hline
Sound100 & \multirow{2}{*}{Audio} & 100MB
& \multirow{2}{=}{ESC dataset for environmental sound~\cite{piczak2015esc}} \\
\cline{1-1} \cline{3-3}
Sound &  & 882MB
&  \\
\hline
Spitzer100 & \multirow{2}{*}{Float} & 100MB 
& \multirow{2}{=}{Spitzer Space Telescope data~\cite{burtscher2008fpc}} \\
\cline{1-1} \cline{3-3}
Spitzer &  & 1.2GB 
&  \\
\hline
AnCa & \multirow{2}{*}{Genome} & 142MB 
& \multirow{2}{=}{DNA sequence corpus~\cite{pratas2018dna}} \\
\cline{1-1} \cline{3-3}
GaGa &  & 148MB 
&  \\
\hline
Rand & Random & 100MB & Random byte stream \\
\hline
\end{tabular}
\label{tab:datasets}
\end{table}

\begin{table*}[t]
\centering
\caption{Compression results in bits per symbol across datasets}
\label{tab:results}
\resizebox{\textwidth}{!}{
\begin{tabular}{|l|ccccccccc|cccccc|c|}
\hline
\textbf{Methods} 
& \textbf{Backup100} 
& \textbf{Enwik8}
& \textbf{Books}
& \textbf{Image100}
& \textbf{Sound100}
& \textbf{Spitzer100}
& \textbf{AnCa}
& \textbf{GaGa} 
& \textbf{Avg}
& \textbf{Backup}
& \textbf{Enwik9}
& \textbf{Image}
& \textbf{Sound}
& \textbf{Spitzer}
& \textbf{Avg}
& \textbf{Rand}\\
\hline
Gzip 
& 7.00 & 2.92 & 2.78 & 6.99 & 5.91 & 7.52 & 2.18 & 2.19 
& 4.69
& 6.24 & 2.58 & 7.00 & 5.85 & 7.53 
& 5.84 & 8.00\\

7z 
& 5.83 & 1.99 & 1.95 & 5.82 & 4.51 & 6.92 & 1.69 & 1.95 
& 3.83
& 5.08 & 1.72 & 5.83 & 4.44 & 7.00 
& 4.81 & 8.00\\

Zstd 
& 6.71 & 2.03 & 1.95 & 6.69 & 5.57 & 7.06 & 1.69 & 1.96 
& 4.21
& 5.65 & 1.72 & 6.65 & 5.44 & 7.08 
& 5.31 & 8.00\\
\hline

TRACE 
& 4.69 & 1.87 & 1.78 & 4.67 & 4.66 & 6.35 & 1.80 & 1.88 
& 3.43
& 4.61 & 1.55 & 4.55 & 4.57 & 6.25 
& 4.31 & 7.98\\

PAC 
& 4.37 & 1.70 & 1.65 & 4.34 & 4.82 & 6.32 & 1.72 & 1.84 
& 3.35
& 4.28 & 1.38 & 4.26 & 4.92 & 6.22 
& 4.21 & 7.98\\

Proposed ($\lambda=0.2$)
& 4.61 & 1.83 & 1.67 & 4.77 & 4.13 & 6.68 & 1.81 & 1.89
& 3.42
& 4.60 & 1.48 & 4.49 & 3.73 & 6.29
& 4.12 & 8.01\\

Proposed ($\lambda=0.01$)
& 4.61 & 1.77 & 1.64 & 4.59 & 3.90 & 6.39 & 1.76 & 1.88
& 3.32
& 4.60 & 1.48 & 4.49 & 3.73 & 6.29
& 4.12 & 8.01\\

\hline
\end{tabular}}
\end{table*}

\begin{table*}[t]
\centering
\small
\caption{Stopping unit selected by the adaptive disabling criterion for different configurations.}
\label{tab:stop_levels}
\resizebox{\textwidth}{!}{%
\begin{tabular}{|c|cccccccc|ccccc|c|}
\hline
\textbf{Configuration} & \textbf{Backup100} & \textbf{Enwik8} & \textbf{Books} & \textbf{Image100} & \textbf{Sound100} & \textbf{Spitzer100} & \textbf{AnCa} & \textbf{GaGa} & \textbf{Backup} & \textbf{Enwik9} & \textbf{Image} & \textbf{Sound} & \textbf{Spitzer} & \textbf{Rand} \\
\hline
Proposed ($\lambda=0.2$) & 6 & 5 & 5 & 5 & 5 & 3 & 3 & 3 & 6 & 6 & 6 & 6 & 6 & 1 \\
\hline
Proposed ($\lambda=0.01$) & 6 & 6 & 6 & 6 & 6 & 6 & 6 & 6 & 6 & 6 & 6 & 6 & 6 & 1 \\
\hline
\end{tabular}}
\end{table*}

\begin{table*}[t]
\centering
\caption{Encoding speed (KB/s) comparison. Reported as X/Y, where X is the total speed and Y is the speed excluding arithmetic encoding operations.}
\label{tab:enc_speed}
\resizebox{\textwidth}{!}{
\begin{tabular}{|l|ccccccccc|cccccc|c|}
\hline
\textbf{Methods}
& \textbf{Backup100}
& \textbf{Enwik8}
& \textbf{Books}
& \textbf{Image100}
& \textbf{Sound100}
& \textbf{Spitzer100}
& \textbf{AnCa}
& \textbf{GaGa}
& \textbf{Avg}
& \textbf{Backup}
& \textbf{Enwik9}
& \textbf{Image}
& \textbf{Sound}
& \textbf{Spitzer}
& \textbf{Avg}
& \textbf{Rand}\\
\hline

TRACE
& 17 / 18
& 17 / 18
& 17 / 18
& 17 / 18
& 16 / 18
& 16 / 18
& 17 / 18
& 17 / 18
& 17 / 18
& 16 / 18
& 17 / 18
& 17 / 18
& 17 / 18
& 16 / 18
& 17 / 18
& 16 / 18 \\

PAC
& 37 / 44
& 39 / 44
& 40 / 45
& 37 / 45
& 36 / 44
& 35 / 45
& 39 / 46
& 39 / 46
& 38 / 45
& 36 / 44
& 39 / 44
& 37 / 44
& 37 / 45
& 35 / 44
& 37 / 44
& 34 / 44 \\

Proposed ($\lambda=0.2$)
& 43 / 57
& 98 / 131
& 124 / 173
& 55 / 82
& 66 / 103
& 83 / 202
& 243 / 814
& 244 / 827
& 120 / 299
& 100 / 221
& 139 / 201
& 103 / 236
& 105 / 227
& 88 / 202
& 107 / 217
& 100 / 894  \\

Proposed ($\lambda=0.01$)
& 43 / 57
& 65 / 78
& 80 / 99
& 36 / 46
& 45 / 59
& 33 / 43
& 108 / 156
& 109 / 158
& 65 / 87
& 100 / 221
& 139 / 201
& 103 / 236
& 105 / 227
& 88 / 202
& 107 / 217 
& 100 / 894 \\

\hline
\end{tabular}}
\end{table*}

\begin{table*}[t]
\centering
\caption{Decoding speed (KB/s) comparison. Reported as X/Y, where X is the total speed and Y is the speed excluding arithmetic decoding operations.}
\label{tab:dec_speed}
\resizebox{\textwidth}{!}{
\begin{tabular}{|l|ccccccccc|cccccc|c|}
\hline
\textbf{Methods}
& \textbf{Backup100}
& \textbf{Enwik8}
& \textbf{Books}
& \textbf{Image100}
& \textbf{Sound100}
& \textbf{Spitzer100}
& \textbf{AnCa}
& \textbf{GaGa}
& \textbf{Avg}
& \textbf{Backup}
& \textbf{Enwik9}
& \textbf{Image}
& \textbf{Sound}
& \textbf{Spitzer}
& \textbf{Avg}
& \textbf{Rand}\\
\hline

TRACE
& 17 / 18
& 16 / 18
& 16 / 18
& 17 / 18
& 17 / 18
& 17 / 18
& 16 / 18
& 16 / 18
& 17 / 18
& 17 / 18
& 16 / 18
& 17 / 18
& 17 / 18
& 17 / 18
& 17 / 18 
& 17 / 18 \\

PAC
& 37 / 43
& 35 / 43
& 35 / 44
& 36 / 43
& 36 / 43
& 36 / 43
& 33 / 43
& 35 / 45
& 35 / 43
& 36 / 43
& 34 / 42
& 35 / 42
& 36 / 44
& 36 / 43
& 35 / 43 
& 36 / 43 \\

Proposed ($\lambda=0.2$)
& 116 / 488
& 259 / 1083
& 314 / 1706
& 127 / 679
& 146 / 759
& 114 / 844
& 422 / 5188
& 429 / 5451
& 241 / 2025
& 121 / 504
& 252 / 818
& 121 / 500
& 141 / 556
& 102 / 430
& 147 / 562 
& 98 / 746 \\

Proposed ($\lambda=0.01$)
& 116 / 488
& 244 / 821
& 294 / 1207
& 121 / 495
& 140 / 560
& 104 / 446
& 373 / 1958
& 382 / 2053
& 222 / 1004
& 121 / 504
& 252 / 818
& 121 / 500
& 141 / 556
& 102 / 430
& 147 / 562 
& 98 / 746 \\

\hline
\end{tabular}}
\end{table*}

Let us denote the weights compression configuration as 
$q = \{\gamma,b,V\}$. Then, the resulting centroid indices are compressed via LZMA2, so that
$R_{\text{w}}(q)$ bits are written into the output bit stream,
while the quantized weights $\widehat{\bm w}_6(q)$
are used for compression of the input data with the neural predictor $\theta_6$, so that approximately 
$\mathcal L_{\mathcal B}(\theta_6,\widehat{\bm w}_6(q))$ bits are spent on batch $\mathcal B$. On the one hand, higher $\gamma$, a lower codebook size $2^b$, or a lower vector size~$V$ reduce the number of bits needed to represent the quantized weights $\widehat{\bm w}_6(q)$. On the other hand, a higher level of quantization of the weights could decrease the compression efficiency of the input data. Therefore, in this paper we search for the best configuration such that
\begin{equation}
q^*=\arg\min_{q\in\mathcal Q}\Big(\sum_{\mathcal B \in \{\mathcal B\}}\mathcal L_{\mathcal B}(\theta_6,\widehat{\bm w}_6(q))+R_{\text{w}}(q)\Big).
\end{equation}

Finally, the token stream is entropy-coded with arithmetic coding~\cite{witten1987arithmetic}. As a result, the output archive bit stream includes:
\begin{enumerate}
\item Arithmetic-encoded token sequence $\bm t_1^{N_t}$.
\item Alphabet $\bm T$ compressed by the LZMA2~\cite{pavlov19997zip} algorithm.
\item Vector quantized weights $\widehat{\bm w}_1,\dots,\widehat{\bm w}_6$ represented as a set of centroid indices compressed by the LZMA2 algorithm. 
\end{enumerate}

\subsection{Adaptive disabling of high-order units}

As was mentioned above, the proposed chained architecture 
allows us to disable high-order units, minimizing the number of weights participating in the probability estimation process.
In this paper, we propose to select the number of units participating
in the probability estimation process taking into account a trade-off
between computational complexity of the encoding and compression efficiency.

Let us introduce the disabling objective as
\begin{equation}
\label{eq:time_aware_stopping}
\mathcal D_i = \mathcal L(\theta_i,\bm w_i) + \lambda \cdot T(\theta_i,\bm w_i),
\end{equation}
where $\mathcal L(\theta_i,\bm w_i)$ is an estimate of the number of bits achieved by unit $i$ after compression, $T(\theta_i,\bm w_i)$ is the encoding time for unit $i$, and $\lambda$ is the Lagrangian multiplier that defines the level of compromise between complexity and compression efficiency.
Then, the proposed disabling procedure is performed as follows. Unit~1, corresponding
to the neural predictor $\theta_1$, is always used. Then, starting from unit~2, the condition $\mathcal D_i<\mathcal D_{i-1}$ 
indicates that an increase in encoding time is accompanied by a significant
increase in compression ratio, i.e., we need to use unit~$i$ as well.
Otherwise, if $\mathcal D_i\geq \mathcal D_{i-1}$, then we use
only units $1,2,...,i-1$ for probability estimation, while the remaining high-order units are disabled.

\begin{figure*}[t]
    \centering
    \begin{tabular}{cccc}
        \begin{subfigure}[b]{0.23\textwidth}
            \includegraphics[width=\textwidth]{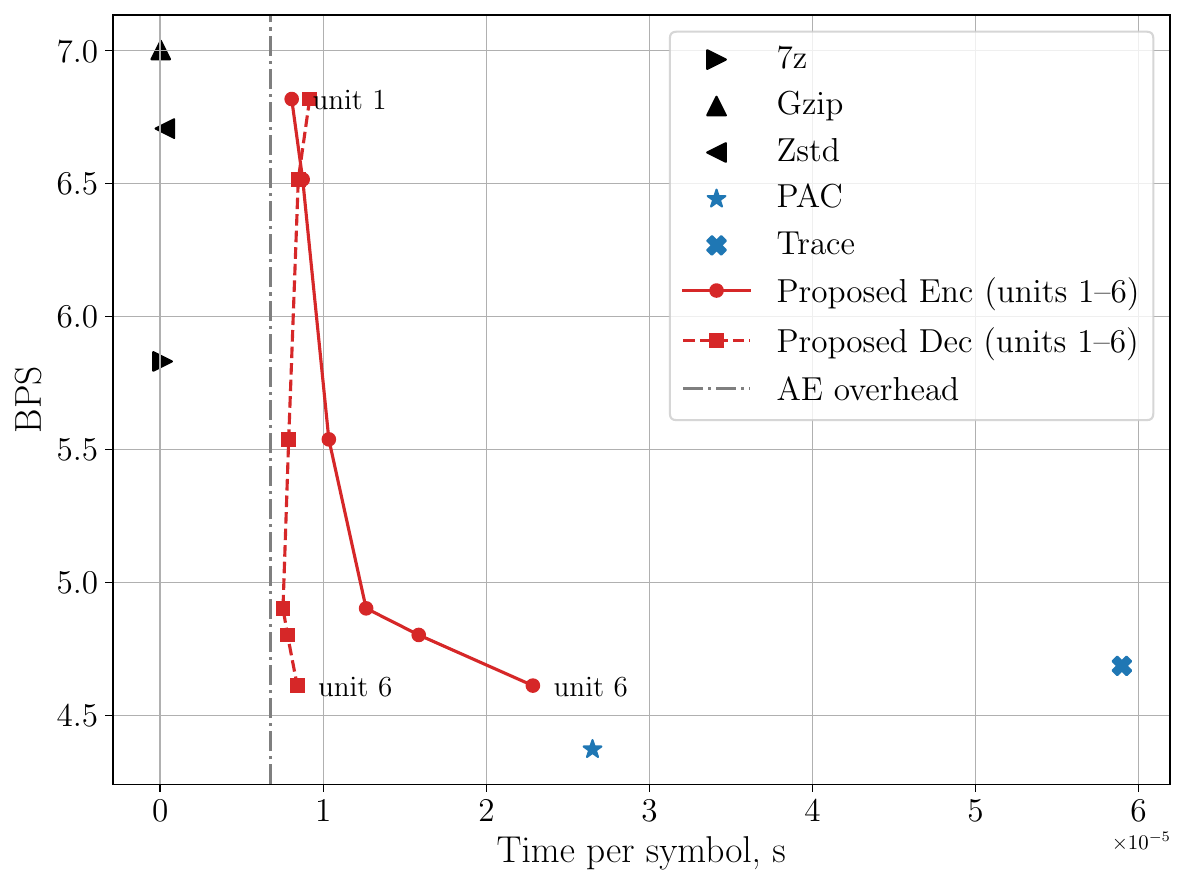}
            \caption{Backup100}
        \end{subfigure} &
        \begin{subfigure}[b]{0.23\textwidth}
            \includegraphics[width=\textwidth]{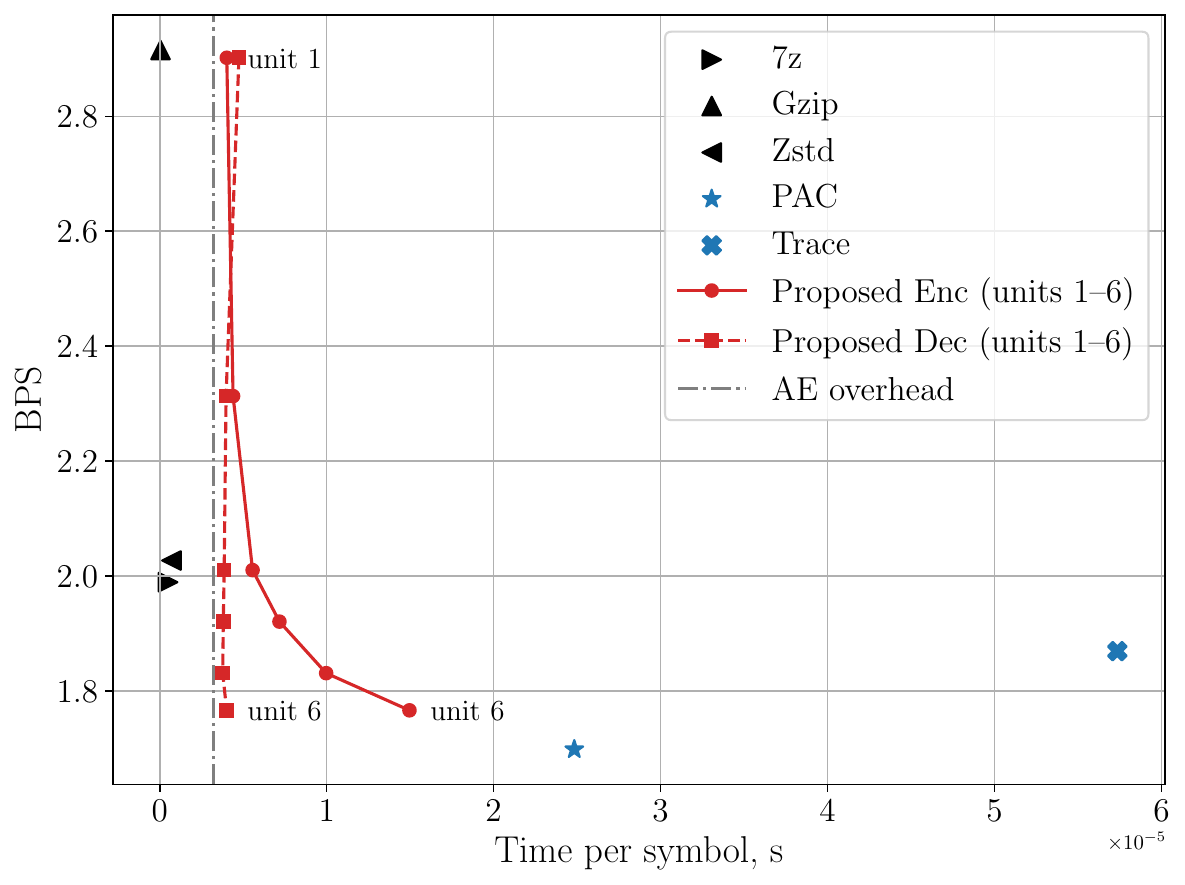}
            \caption{Enwik8}
        \end{subfigure} &
        \begin{subfigure}[b]{0.23\textwidth}
            \includegraphics[width=\textwidth]{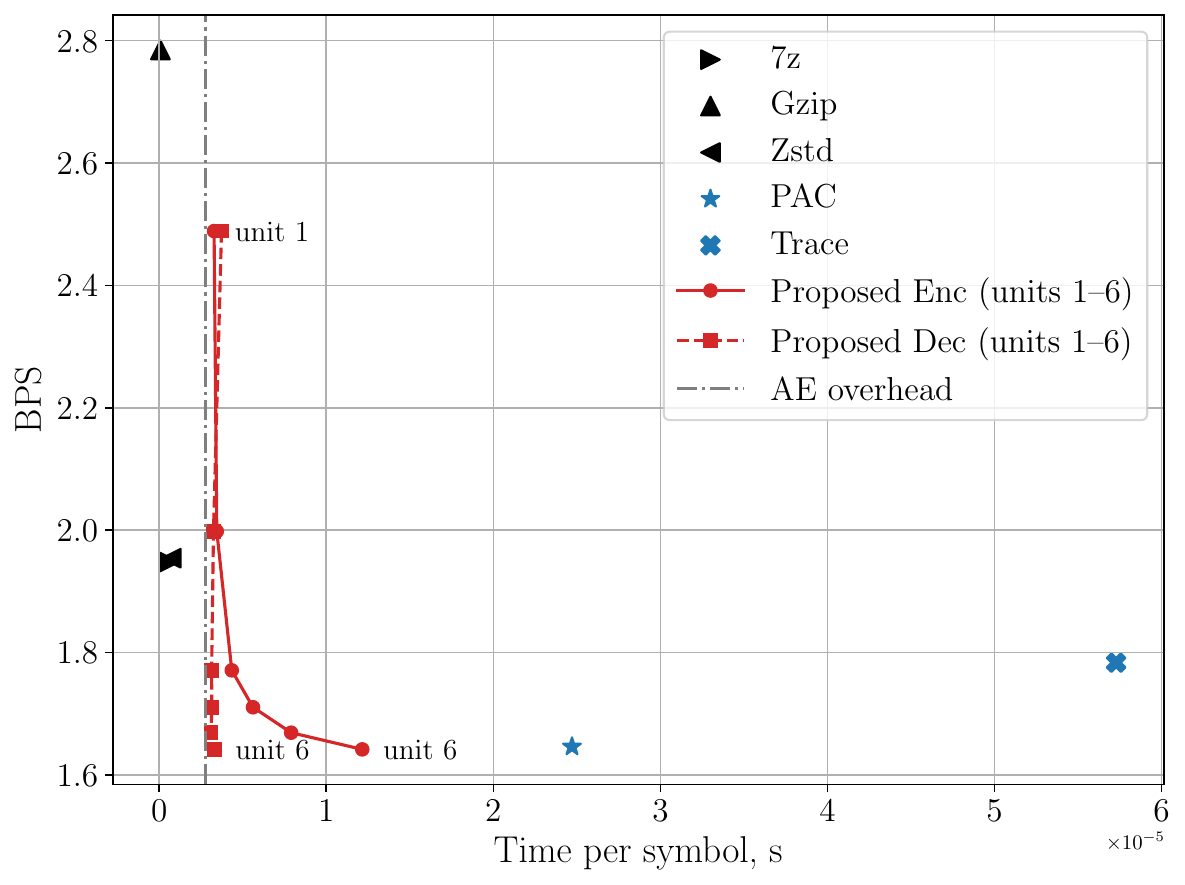}
            \caption{Books}
        \end{subfigure} &
        \begin{subfigure}[b]{0.23\textwidth}
            \includegraphics[width=\textwidth]{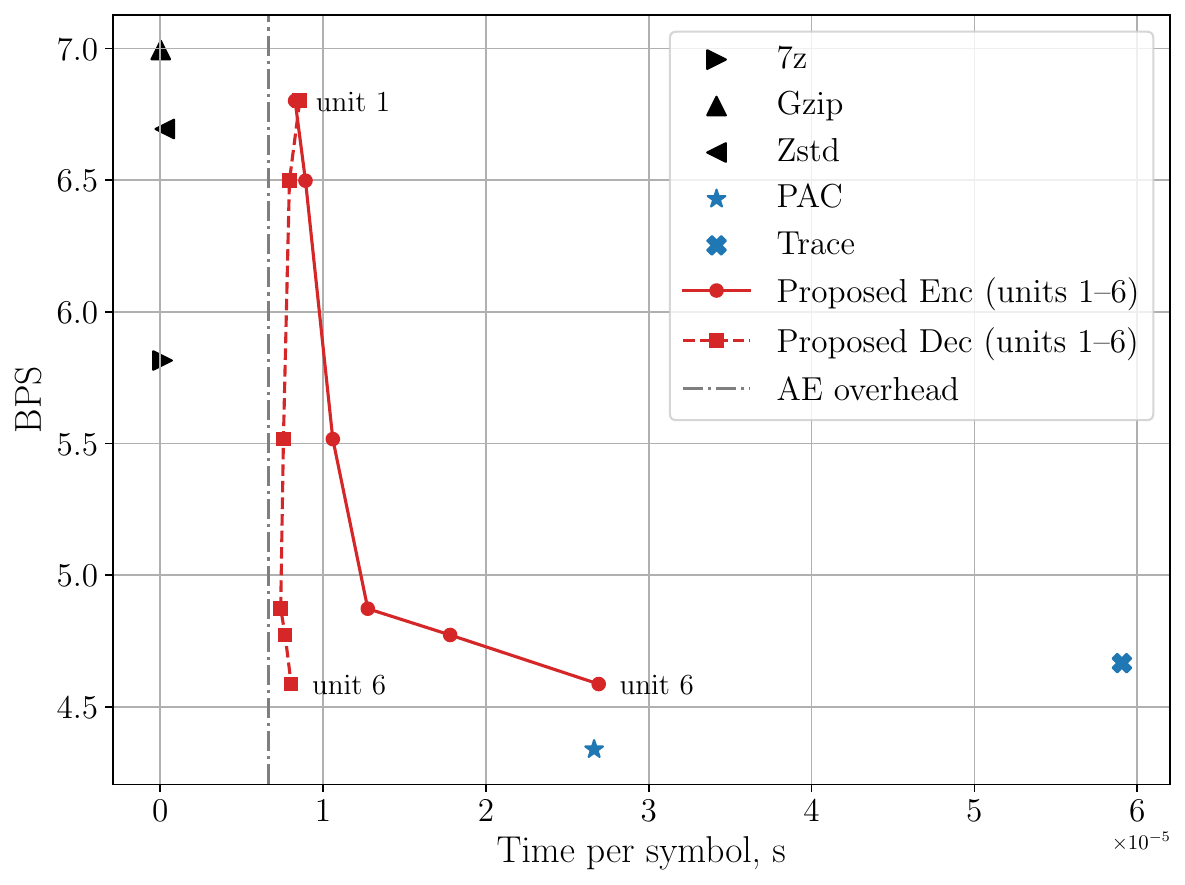}
            \caption{Image100}
        \end{subfigure} \\
        \begin{subfigure}[b]{0.23\textwidth}
            \includegraphics[width=\textwidth]{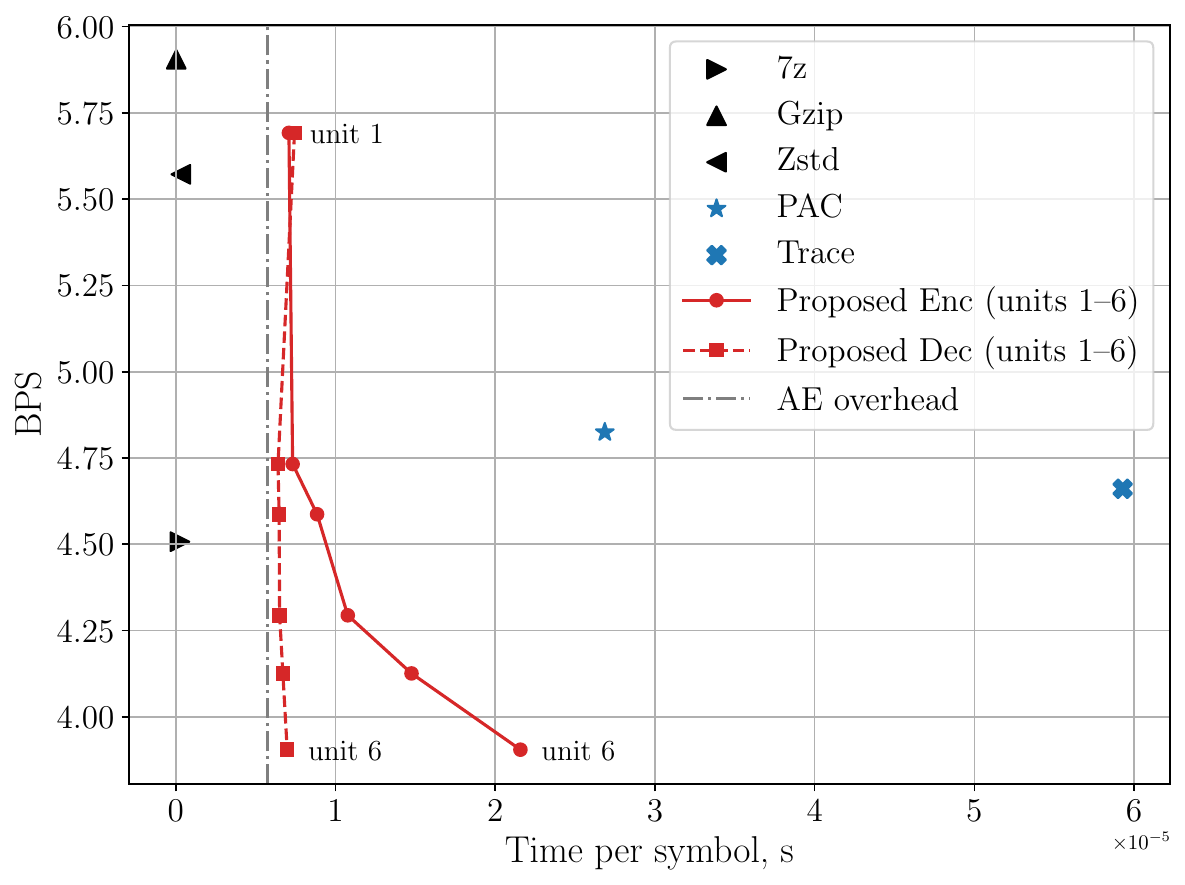}
            \caption{Sound100}
        \end{subfigure} &
        \begin{subfigure}[b]{0.23\textwidth}
            \includegraphics[width=\textwidth]{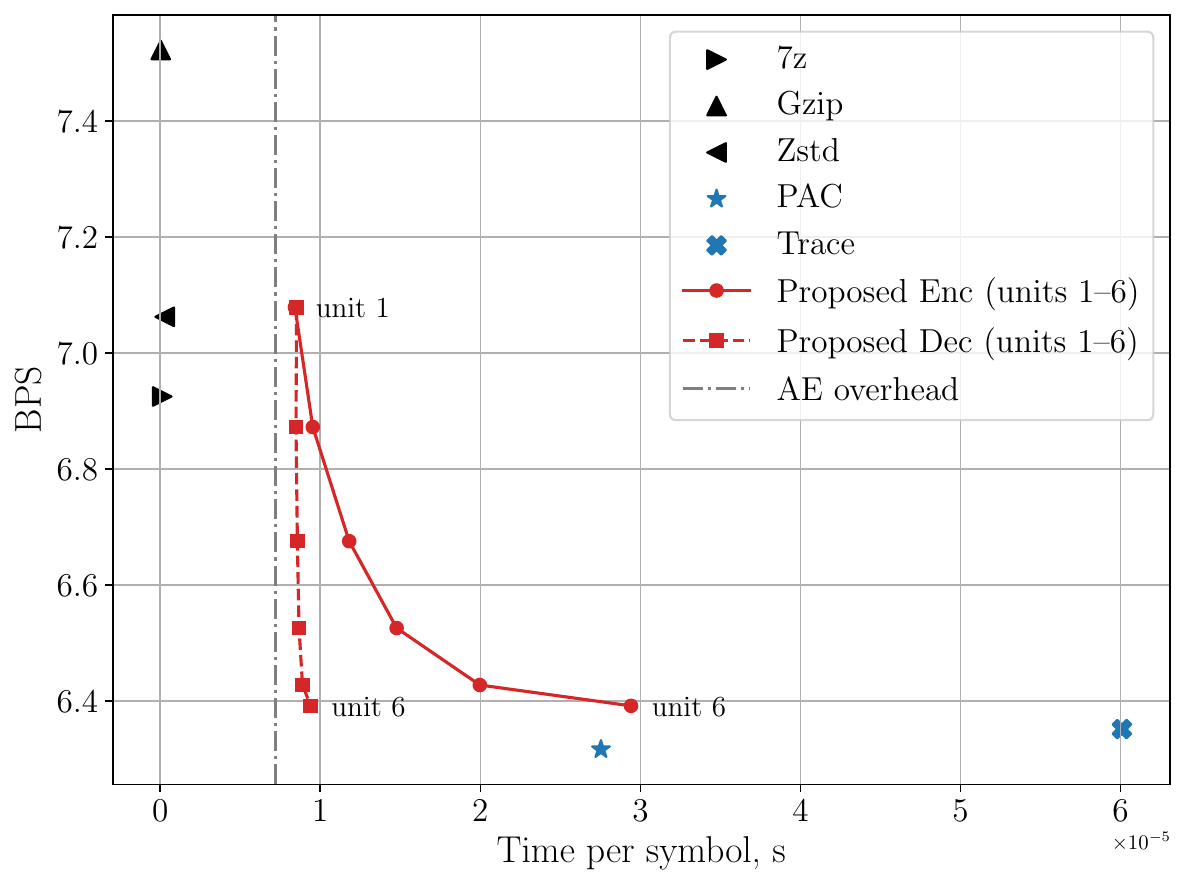}
            \caption{Spitzer100}
        \end{subfigure} &
        \begin{subfigure}[b]{0.23\textwidth}
            \includegraphics[width=\textwidth]{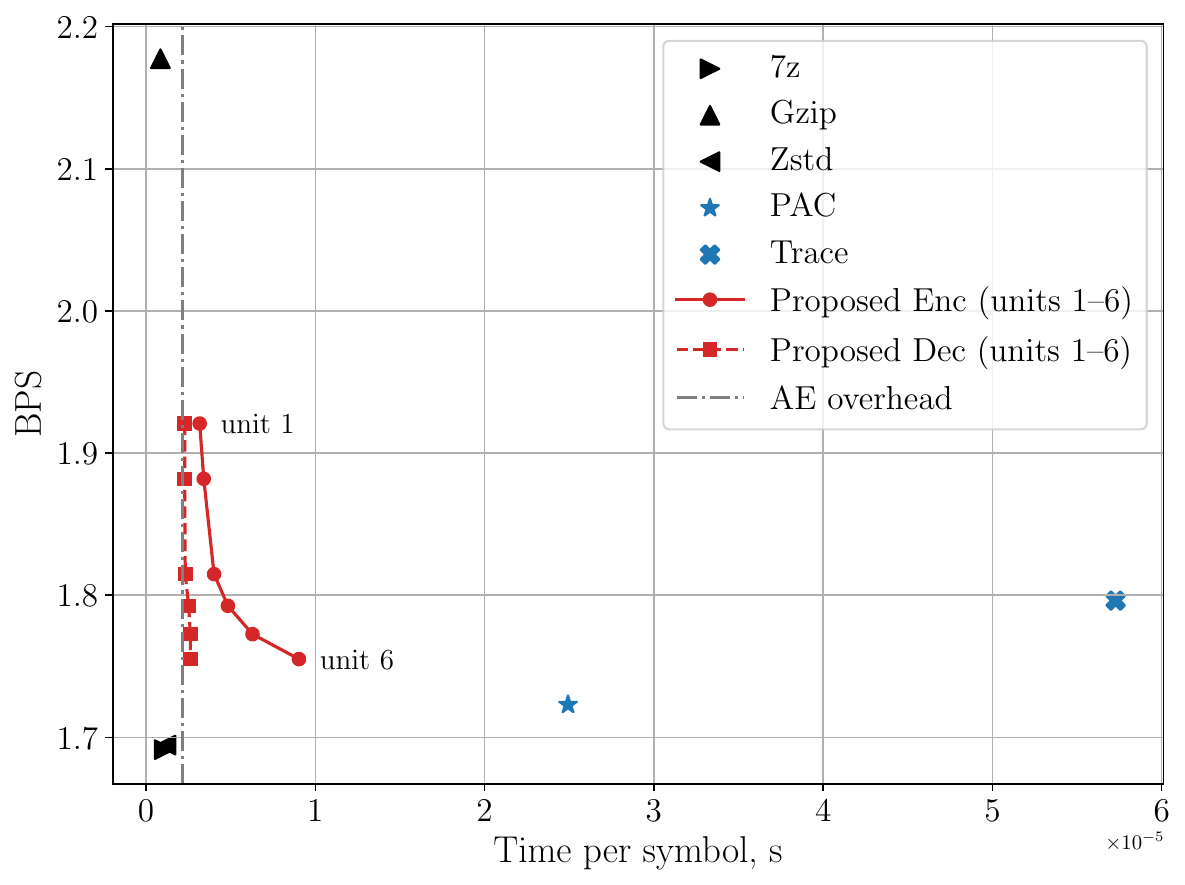}
            \caption{AnCa}
        \end{subfigure} &
        \begin{subfigure}[b]{0.23\textwidth}
            \includegraphics[width=\textwidth]{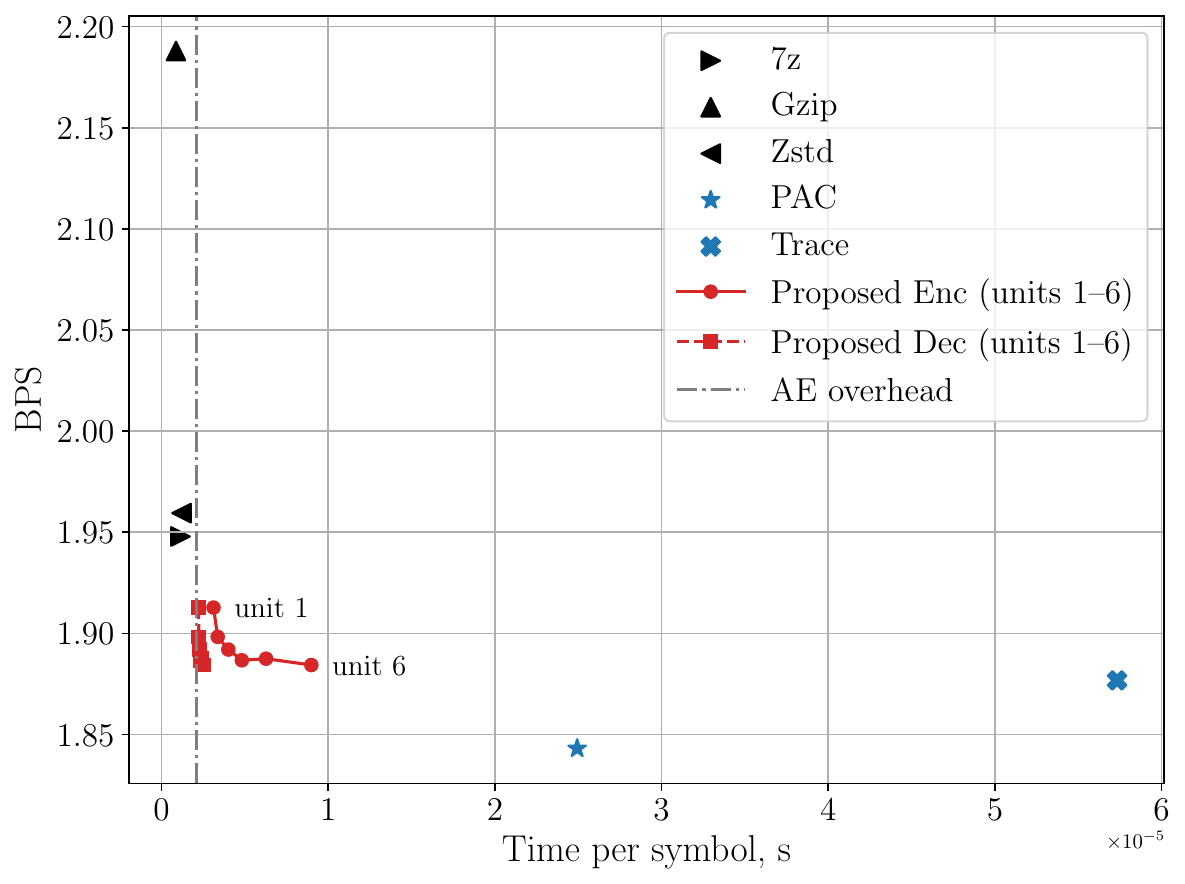}
            \caption{GaGa}
        \end{subfigure}
    \end{tabular}
    \caption{\label{fig:bpp_vs_time} Compression efficiency versus processing time comparison for different compressors}
\end{figure*}

\begin{table}[t]
\centering
\caption{Peak GPU memory, number of parameters, and inference FLOPs of individual neural-based models with batch 512. GPU memory is measured over the full pipeline, including training and inference.}
\label{tab:model_complexity}
\resizebox{\columnwidth}{!}{
\begin{tabular}{|l|cc|cccccc|}
\hline
\textbf{Metric}
& \textbf{TRACE}
& \textbf{PAC}
& \textbf{Unit 1}
& \textbf{Unit 2}
& \textbf{Unit 3}
& \textbf{Unit 4}
& \textbf{Unit 5}
& \textbf{Unit 6}\\
\hline
GPU Memory (GB) & 0.85 & 0.39 & 0.32 & 0.37 & 0.37 & 0.44 & 0.43 & 0.77 \\
Parameters (M) & 2.37 & 19.74 & 0.02 & 0.14 & 0.76 & 0.55 & 0.79 & 0.82 \\
FLOPs (G) & 36.58 & 8.68 & 0.01 & 0.12 & 0.70 & 0.28 & 0.56 & 0.56 \\
\hline
\end{tabular}}
\end{table}

\section{Experiments}
\label{sec:experiments}
\subsection{Experimental Setup}
\label{sec:setup}

We evaluated the proposed method on seven datasets with different modalities, summarized in Table~\ref{tab:datasets}. For several datasets, we additionally used truncated versions in order to analyze the behavior of the method at different data scales. All experiments were conducted on a system equipped with an Intel Core i5-12400F CPU, an NVIDIA GeForce RTX 4060 Ti GPU, and 64 GB of DDR4 RAM. 

We used gzip~\cite{deutsch1996gzip}, zstd~\cite{collet2018zstandard}, and 7z~\cite{pavlov19997zip}, all configured at their maximum compression levels (gzip -9, zstd -22, and 7z -mx=9) as classical baselines. For neural baselines we used the PAC~\cite{mao2023pac} and TRACE~\cite{mao2022trace} compressors with default parameters. Arithmetic coding follows the configuration described in~\cite{mao2023pac}.

For the proposed compressor the following hyperparameters were used: the batch size was set to $|\mathcal B|=4096$, and each unit was trained for $E_1=1$ epoch without information inheritance, followed by $E_2=2$ epochs with information inheritance enabled. Optimization was performed using the Adam optimizer with an initial learning rate of $4 \times 10^{-3}$. The learning rate was decayed by a factor of $0.8$ after each epoch. Early stopping was applied based on the validation loss, evaluated every $K_{\mathrm{v}} = 1000$ iterations, with patience $P = 5$. The validation coefficient was fixed to $\rho = 10^{-3}$. For model weights compression, the configuration search was carried out over $b \in \{4, 8\}$, $\gamma \in \{1, 2, \dots, 9\} \times 10^{-5}$, and $V \in \{1, 2, 4\}$.

\subsection{Compression Performance}
Table~\ref{tab:results} presents the compression performance of the proposed compressor compared with other methods. Compression efficiency is evaluated in terms of BPS, defined as the total number of encoded bits divided by the number of input symbols.
One can see that the classical compressors (Gzip, 7z, and Zstd) are generally less competitive than neural approaches, confirming the advantage of learned probabilistic modeling across heterogeneous domains. On the smaller datasets, PAC achieves 3.35~BPS on average, while the proposed compressor with $\lambda=0.2$ (the balanced configuration between compression efficiency and complexity) reaches 3.42~BPS. Thus, the average BPS of the balanced-oriented setting is only slightly higher. When $\lambda$ is reduced to 0.01, the gap becomes smaller: the proposed compressor reaches 3.32~BPS on average, which is very close to the 3.35~BPS achieved by PAC. On the larger datasets, both proposed configurations achieve 4.12~BPS on average, compared with 4.21~BPS for PAC and 4.31~BPS for TRACE.

The adaptive disabling results in Table~\ref{tab:stop_levels} 
explain the difference in compression efficiency for the proposed compressor with $\lambda=0.2$
and $\lambda=0.01$. For several smaller datasets, fewer than six units are sufficient: for example, the proposed compressor with $\lambda=0.2$ stops at Unit~5 on Enwik8, Books, Image100, and Sound100, and even at Unit~3 on Spitzer100, AnCa, and GaGa. In contrast, the compression-oriented configuration with $\lambda=0.01$ more often uses all six units, reflecting its stronger emphasis on compression performance. For the large datasets, both configurations always reach Unit~6. This is consistent with the fact that, relative to the dataset size, the corresponding model sizes remain small, so using the full chain is generally worthwhile. On the random dataset, however, both configurations stop already at Unit~1, which is expected because random data contain no meaningful statistical dependencies that could justify training higher-order units.

\subsection{Computational Efficiency}
Table~\ref{tab:enc_speed} presents the encoding speed of the evaluated neural compressors, measured in kilobytes per second (KB/s). The reported speed is calculated by dividing the size of the input file by the total runtime of the compressor, which includes model training, arithmetic coding, and all other preprocessing operations.

Across all datasets, the proposed method consistently achieves the highest encoding throughput, clearly outperforming both TRACE and PAC. Relative to PAC, the balanced-oriented configuration Proposed~($\lambda=0.2$) provides encoding speedups ranging from about 1.2$\times$ on Backup100 (43 vs. 37~KB/s) to about 6.3$\times$ on GaGa (244 vs. 39~KB/s). At the same time, excluding arithmetic coding, Proposed~($\lambda=0.2$) provides encoding speedups over PAC ranging from about 1.3$\times$ on Backup100 (57 vs. 44~KB/s) to about 20.3$\times$ on Rand (894 vs. 44~KB/s), which indicates that the advantage is not limited to the entropy-coding stage and largely comes from more efficient probability estimation.
These results show that the proposed complexity-adaptive unit disabling approach effectively reduces unnecessary computation during encoding, particularly for sources that do not require uniformly high model capacity at every stage.

Table~\ref{tab:dec_speed} reports the decoding speed of the compressors. The decoding time accounts for all operations required to reconstruct the original data, providing a direct measure of each method's runtime efficiency.
Compared with TRACE and PAC, the proposed method delivers the strongest decoding-speed advantage across all data domains. Relative to PAC, Proposed~($\lambda=0.2$) achieves decoding speedups ranging from about 2.8$\times$ on Spitzer (102 vs. 36~KB/s) to about 12.3$\times$ on GaGa (429 vs. 35~KB/s). Excluding arithmetic decoding, Proposed~($\lambda=0.2$) provides decoding speedups over PAC ranging from about 10.0$\times$ on Spitzer (430 vs. 43~KB/s) to about 121.1$\times$ on GaGa (5451 vs. 45~KB/s). This behavior is consistent with the design objective of the method: to minimize repeated adaptive computation at reconstruction time. From a practical perspective, these results are important because compressed content is often decoded multiple times, so decoding efficiency has a strong impact on end-to-end deployment cost.

Table~\ref{tab:model_complexity} complements these runtime results by showing that the units in the proposed chain remain substantially smaller than PAC in terms of the number of parameters and FLOPs, which explains why adaptive unit selection can translate into large practical speedups.

\subsection{Ablation Study}
To better understand the behavior of the proposed method, we perform an ablation study analyzing the influence of several key design parameters. 
Unless otherwise stated, all experiments use the default configuration described in Section~\ref{sec:setup}, while varying one parameter at a time.

\subsubsection{Effect of Information Inheritance}

\begin{table}[t]
\centering
\caption{BPS reduction (in \%) achieved by the information inheritance between units compared with a single unit}
\label{tab:ablation_ii}
\resizebox{\columnwidth}{!}{
\begin{tabular}{|l|cccccc|}
\hline
\textbf{Dataset}
& \textbf{Unit 1}
& \textbf{Unit 2}
& \textbf{Unit 3}
& \textbf{Unit 4}
& \textbf{Unit 5}
& \textbf{Unit 6}\\
\hline
Backup & 0.0 & 0.4 & 1.3 & 1.9 & 2.5 & 2.3 \\
Enwik9 & 0.0 & 1.6 & 2.6 & 6.6 & 7.0 & 8.6 \\
Image & 0.0 & 0.26 & 0.76 & 0.74 & 1.79 & 1.42 \\
Sound & 0.0 & 0.2 & 0.6 & 0.5 & 0.8 & 1.7 \\
Spitzer & 0.0 & 0.1 & 0.6 & 0.6 & 1.2 & 13.9 \\
\hline
\end{tabular}}
\end{table}

Table~\ref{tab:ablation_ii} reports the relative BPS reduction obtained by enabling information inheritance, measured with respect to standalone models trained independently (with $\alpha_i=1$, $\beta_i=0$). As expected, no reduction is observed at Unit~1, since there is no lower-order predictor from which information can be inherited. From Unit~2 onward, the relative BPS reduction generally increases with the unit index. In particular, on Enwik9 the reduction reaches 6.6\% at Unit~4 and further increases to 8.6\% at Unit~6. An even stronger effect is observed on Spitzer, where the reduction remains modest up to Unit~5 but rises sharply to 13.9\% at Unit~6.

\subsubsection{Validation size}

A smaller validation subset is desirable because it reduces the overhead of early stopping and model-selection procedures, which is especially important in the semi-adaptive setting. To study this trade-off, we vary the validation coefficient $\rho$, which controls the subset size.
Table~\ref{tab:val_size_ablation} shows that the estimated BPS changes only slightly when the validation coefficient varies over a wide range. Therefore, using $\rho=10^{-3}$, i.e., validating on 0.1\% of the input token sequence, provides an estimate that is sufficiently close to those obtained with larger validation subsets, while keeping the corresponding computational overhead low.

\begin{table}[t]
\centering
\caption{Validation coefficient vs. estimated compression performance on the Backup100 dataset}
\label{tab:val_size_ablation}
\resizebox{\columnwidth}{!}{
\begin{tabular}{|c|c|c|c|c|c|c|}
\hline
$\rho$ & $10^{-3}$ & $5\times10^{-3}$ & $10^{-2}$ & $5\times10^{-2}$ & $10^{-1}$ & $1$ \\
\hline
BPS & 4.727 & 4.726 & 4.726 & 4.730 & 4.730 & 4.731 \\
\hline
\end{tabular}}
\end{table}

\section{Conclusion}
\label{sec:conclusion}
In this paper, we proposed a lossless compression approach that combines a complexity-adaptive chain of neural predictors with information inheritance to improve prediction efficiency. We evaluated its performance across different data modalities and dataset sizes and showed substantial encoding-speed gains, multiple-fold decoding-speed gains, and a minor improvement in average compression ratio. Therefore, the proposed approach provides a promising step toward computationally efficient neural lossless data compression.

\bibliographystyle{IEEEtran}
\bibliography{ref}
\begin{IEEEbiographynophoto}{Yuriy Kim} is a PhD student
at ITMO University, Saint-Petersburg, Russia. His current research interests include neural data and image compression. He received a Master's degree in 2024 from ITMO University, Saint-Petersburg, Russia. Contact him at ylkim@itmo.ru.
\end{IEEEbiographynophoto}
\begin{IEEEbiographynophoto}{Evgeny Belyaev} is an Associate Professor
at ITMO University, Saint-Petersburg, Russia. He received the Candidate of Science degree in 2009 from the State University of Aerospace Instrumentation (SUAI), Saint-Petersburg, Russia, and the Doctor of Technology degree from Tampere University of Technology, Finland, in 2015. His current research interests include image and video compression, source coding, and compressive sensing. Contact him at eabelyaev@itmo.ru.
\end{IEEEbiographynophoto}
\end{document}